\documentstyle[psfig]{l-aa}

\begin{document}

\thesaurus{ 08.01.2, 08.02.1, 08.03.3, 08.06.1, 08.12.1 }

\title {Multiwavelength optical observations of chromospherically active binary
systems}
\subtitle{I. Simultaneous H$\alpha$, Na~{\sc i} D$_{1}$, D$_{2}$,
and He~{\sc i} D$_{3}$ observations 
\thanks{Based on 
observations made with the Isaac Newton telescope
operated on the island of La Palma by the
Royal Greenwich Observatory at
the Spanish Observatorio del Roque de Los Muchachos of the
 Instituto de Astrof\'{\i}sica de Canarias
} }

\author{
D.~Montes
\and M.J.~Fern\'{a}ndez-Figueroa
\and E.~De Castro
\and J.~Sanz-Forcada
}

\offprints{ D.~Montes}

\institute{Departamento de Astrof\'{\i}sica,
Facultad de F\'{\i}sicas,
 Universidad Complutense de Madrid, E-28040 Madrid, Spain
\\  E-mail: dmg@ucmast.fis.ucm.es}

\date{Received ; accepted }

\maketitle
\markboth{D. Montes et al.: Simultaneous H$\alpha$, 
Na~{\sc i} D$_{1}$, D$_{2}$, and He~{\sc i} D$_{3}$ observations}{ }

\begin{abstract}

This is the first paper of a series aimed at studying
the chromosphere of active binary systems
using the information provided for several optical spectroscopic 
features.
Simultaneous H$\alpha$, Na~{\sc i} D$_{1}$, D$_{2}$,
and He~{\sc i} D$_{3}$ spectroscopic observations are reported here
for 18 systems.
The chromospheric contribution in these lines have been determined
using the spectral subtraction technique.
Very broad wings have been found in the subtracted H$\alpha$ profile
of some of the more active stars.
These profiles are well matched using a two-components
Gaussian fit (narrow and broad) and the broad component 
could be interpreted as arising from microflaring.
Prominence-like extended material have been detected in a near-eclipse
H$\alpha$ observation of the system AR Lac.
The excess emission found in the Na~{\sc i} D$_{1}$ and D$_{2}$ lines
by application of the spectral subtraction technique and the
behaviour of the H$\alpha$ line in the corresponding simultaneous 
observations indicate that the filling-in of the core of these lines
is a chromospheric activity indicator.
For giant stars of the sample
the He~{\sc i} D$_{3}$ line has been detected in absorption
in the subtracted spectra.
An optical flare has been detected in UX Ari and II Peg through
the presence of the He~{\sc i} D$_{3}$ in emission in coincidence with the
enhancement of the H$\alpha$ emission.
%
%
\keywords{  stars: activity  -- stars: binaries: close
 -- stars: chromospheres -- stars: flare -- stars: late-type
}

\end{abstract}


\section{Introduction}

The chromospherically active binaries are
detached binary systems with cool components
characterized by strong chromospheric, transition region, and coronal
activity. The RS CVn systems have at least one cool evolved component
whereas both components of the BY Dra  binaries are main sequence stars 
(Fekel et al. 1986).

In this series of papers we try to study
the chromosphere of this kind of extremely active stars 
using the information provided by several optical spectroscopic 
features that could be used as chromospheric activity indicators.
The simultaneous observations of different lines,
that are formed at different height in the chromosphere
(from the region of temperature minimum to the higher chromosphere), 
are of special interest for stellar activity studies since they 
provide very useful information about this 
stellar region.
Ideally, simultaneous observations should be performed at all wavelengths in 
order to develop a coherent 3-D atmosphere model.
In practice, simultaneous observations of several activity indicators are 
rare and tend to focus on the same small number of extremely active systems.

The best way to obtain the active-chromosphere contribution to 
some spectral line in the chromospherically active binaries
is to subtract the underlying photospheric
contribution using the spectral subtraction technique
(subtraction of a synthesized stellar spectrum constructed
from artificially rotationally broadened, radial-velocity
shifted, and weighted spectra of inactive stars
chosen to match the spectral types and luminosity classes
of both components of the active
system under consideration)

The emissions in the Ca~{\sc ii} H \& K resonance lines are the
most widely used optical indicators of chromospheric activity, 
since their source functions are collisionally controlled and 
represent an extremely important cooling mechanism.
In chromospheric active binaries  
the subtraction of the photospheric flux in this spectral region
has been recently applied
using the spectral subtraction
(see Montes et al. 1995c, 1996a and references therein).

The H$\alpha$ line is also an important chromospheric activity
indicator, but it is only in emission above the continuum in very active stars,
and in less active star only a filled-in absorption line is observed. 
So, to infer chromospheric activity level the spectral subtraction is needed
(see Montes et al. 1994; 1995a; b; d, and references therein; 
L\'{a}zaro \& Ar\'{e}valo 1996). 
A similar behaviour is observed in the other Balmer lines
(Hall \& Ramsey 1992; Montes et al. 1995d).

Recently, the spectral subtraction technique has been used 
in other lines as the
Ca~{\sc ii} IRT, Mg~{\sc i} b, Na~{\sc i} D$_{1}$, D$_{2}$, 
and He~{\sc i} D$_{3}$ lines (Gunn \& Doyle 1996; Gunn et al. 1996). 
The Ca~{\sc ii} IRT lines  are formed deeper in the atmosphere
and are thus sensitive probes of the temperature minimum region.
The Na~{\sc i} D$_{1}$, D$_{2}$ lines are collision dominated and are good
indicators of changes in the lower chromosphere.
The Mg~{\sc i} b triplet lines are formed in the lower chromosphere and the
region of temperature minimum and they are good diagnostics of 
photospheric activity (Basri et al. 1989).
The He~{\sc i} D$_{3}$ line has been largely ignored as activity indicator;
however it could be a valuable probe of stellar activity and the observation
of this line in emission supports the detection of flare like events
(Zirin 1988).

In this first paper we focus 
our study on the analysis of the extensively used H$\alpha$ 
chromospheric activity indicator together with simultaneous
observations of the less studied He~{\sc i} D$_{3}$ and 
Na~{\sc i} D$_{1}$, D$_{2}$ spectral features
in a sample of 18 northern active binary systems
selected from "A Catalog of Chromospherically Active Binary Stars
(second edition)" (Strassmeier et al. 1993, hereafter CABS). 
By using the spectral subtraction technique,
we have determined the excess emission in these lines and 
we have computed absolute chromospheric fluxes in H$\alpha$.
The primary aim of this study is analyse in detail the 
excess H$\alpha$ emission 
and to study the subtracted H$\alpha$ line profile, especially in
some extremely active stars which exhibit broad wings.
Moreover, we try to understand the behaviour of the
He~{\sc i} D$_{3}$ and Na~{\sc i} D$_{1}$, D$_{2}$ lines as 
chromospheric activity indicators taking into account the advantage
that we simultaneously know the behaviour of the 
chromospheric excess H$\alpha$ emission in these systems.
In forthcoming papers we will analyze in detail several 
optical spectroscopic features using echelle spectroscopy
in order to determine the effects of stellar activity on spectral
lines originating at different heights in the chromosphere.
Another of our goals is to obtain information about the presence of extended
matter (prominence-like structures) in the chromospheric active binaries 
using simultaneous H$\alpha$ and H$\beta$ observations 
at near-eclipse orbital phases.


In Sect.~2 we give the details of our observations and data reduction.
In Sect.~3 we describe the individual results of 
H$\alpha$, Na~{\sc i} D$_{1}$, D$_{2}$, and He~{\sc i} D$_{3}$
line observations of our sample. 
Finally in Sect.~4 we discuss our results.

\begin{table*}
\caption[]{H$\alpha$ and Na~{\sc i} D$_{1}$, D$_{2}$, He~{\sc i} D$_{3}$  
Observing log, and previous H$\alpha$ and Ca~{\sc ii} H \& K observations
\label{tab:obslog}}
\begin{flushleft}
\scriptsize
\begin{tabular}{lcccccccccccc}
\noalign{\smallskip}
\hline
\noalign{\smallskip}
Name &
\multicolumn{4}{c}{H$\alpha$ line region} &\ &
\multicolumn{4}{c}{Na~I D$_{1}$, D$_{2}$, 
He~I D$_{3}$ line region} &\ &
\multicolumn{2}{c}{Previous obs.} \\
\cline{2-5}\cline{7-10}\cline{12-13}
\noalign{\smallskip}
     &
 {Date} & UT & {$\varphi$} & S/N &\ &
 {Date} & UT & {$\varphi$} & S/N &\ &
H$\alpha$ & Ca II \\
\noalign{\smallskip}
\hline
\noalign{\smallskip}
BD Cet    & 1995/09/15 & 00:43 & 0.569 & 330 & & 1995/09/15 & 01:03 & 0.569 & 361 & & -  & 95c,96a \\
\noalign{\smallskip}
AY Cet    & 1995/09/15 & 02:03 & 0.797 & 385 & & 1995/09/15 & 02:20 & 0.797 & 398 & & -  & 95c,96a \\
\noalign{\smallskip}
AR Psc    & 1995/09/13 & 01:48 & 0.373 & 387 & & 1995/09/13 & 01:27 & 0.372 & 141 & & 94,95a  & 94,96a \\
"         & 1995/09/14 & 01:42 & 0.443 & 361 & & 1995/09/14 & 02:00 & 0.443 & 369 & & -  & -  \\
"         & 1995/09/15 & 03:55 & 0.519 & 392 & & 1995/09/15 & 04:23 & 0.520 & 399 & & -  & -  \\
"         & 1995/09/15 & 05:34 & 0.524 & 305 & & 1995/09/15 & 05:28 & 0.523 & 359 & & -  & -  \\
\noalign{\smallskip}
HD 12545  & 1995/09/15 & 02:59 & 0.401 & 354 & & 1995/09/15 & 02:25 & 0.400 & 361 & & -  & 95c,,96a \\
\noalign{\smallskip}
UX Ari    & 1995/09/13 & 02:08 & 0.419 & 346 & &            &       &       &     & & 95b & 95c,96a \\
"         & 1995/09/13 & 05:08 & 0.438 & 202 & & 1995/09/13 & 04:52 & 0.437 & 494 & & -  & -  \\ 
"         & 1995/09/14 & 02:24 & 0.576 & 276 & & 1995/09/14 & 02:09 & 0.574 & 293 & & -  & -  \\
"         & 1995/09/15 & 03:09 & 0.736 & 352 & & 1995/09/15 & 03:32 & 0.739 & 410 & & -  & -  \\
\noalign{\smallskip}
V711 Tau  & 1995/09/13 & 05:15 & 0.922 & 360 & & 1995/09/13 & 05:36 & 0.927 & 370 & & 94,95a & 94,96a \\
"         & 1995/09/14 & 04:19 & 0.261 & 374 & & 1995/09/14 & 04:35 & 0.265 & 423 & & -  & -  \\
"         & 1995/09/14 & 05:35 & 0.280 & 332 & & 1995/09/14 & 05:50 & 0.283 & 386 & & -  & -  \\
"         & 1995/09/15 & 03:50 & 0.606 & 342 & & 1995/09/15 & 03:44 & 0.605 & 332 & & -  & -  \\
"         & 1995/09/15 & 06:11 & 0.641 & 314 & & 1995/09/15 & 06:05 & 0.639 & 326 & & -  & -  \\
\noalign{\smallskip}
V833 Tau  & 1995/09/13 & 06:13 & 0.762 & 184 & & 1995/09/15 & 05:47 & 0.752 & 325 & & 95b & -  \\
"         & 1995/09/14 & 06:07 & 0.319 & 424 & & 1995/09/14 & 05:55 & 0.314 & 417 & & -  & -  \\
"         & 1995/09/15 & 04:56 & 0.851 & 394 & & 1995/09/15 & 05:18 & 0.859 & 322 & & -  & -  \\
\noalign{\smallskip}
V1149 Ori & 1995/09/15 & 06:17 & 0.439 & 311 & & 1995/09/15 & 06:29 & 0.439 & 335 & & 95b & 95c,96a \\
\noalign{\smallskip}
MM Her    & 1995/09/12 & 20:56 & 0.498 & 274 & & 1995/09/12 & 21:31 & 0.501 & 329 & & 94,95a & 94,96a \\
"         & 1995/09/13 & 22:06 & 0.630 & 332 & & 1995/09/13 & 20:45 & 0.623 & 384 & & -  & -  \\
"         & 1995/09/14 & 20:09 & 0.745 & 257 & & 1995/09/14 & 20:41 & 0.748 & 314 & & -  & -  \\
\noalign{\smallskip}
V815 Her  & 1995/09/12 & 20:43 & 0.978 & 360 & & 1995/09/12 & 20:21 & 0.970 & 341 & & 94,95a & 94,96a \\
"         & 1995/09/13 & 20:15 & 0.520 & 375 & & 1995/09/13 & 20:22 & 0.523 & 341 & & -  & -  \\
"         & 1995/09/14 & 21:24 & 0.099 & 354 & & 1995/09/14 & 21:44 & 0.107 & 394 & & -  & -  \\ 
\noalign{\smallskip}
BY Dra    & 1995/09/13 & 22:58 & 0.684 & 387 & & 1995/09/13 & 23:15 & 0.686 & 372 & & -  & 94,96a \\
"         & 1995/09/14 & 21:56 & 0.839 & 398 & & 1995/09/14 & 21:05 & 0.838 & 404 & & -  & -  \\
\noalign{\smallskip}
V775 Her  & 1995/09/14 & 22:03 & 0.394 & 342 & & 1995/09/14 & 21:53 & 0.392 & 373 & & 94,95a & 94,96a  \\
\noalign{\smallskip}
V478 Lyr  & 1995/09/14 & 22:14 & 0.953 & 326 & & 1995/09/14 & 22:33 & 0.959 & 412 & & -  & 94,96a  \\ 
\noalign{\smallskip}
HK Lac    & 1995/09/12 & 22:56 & 0.067 & 343 & & 1995/09/12 & 22:42 & 0.067 & 357 & & 94,95a & 94,96a  \\
"         & 1995/09/13 & 00:42 & 0.070 & 130 & & 1995/09/13 & 00:25 & 0.070 & 193 & & -  & -  \\
"         & 1995/09/13 & 23:44 & 0.110 & 346 & & 1995/09/13 & 23:58 & 0.110 & 343 & & -  & -  \\
"         & 1995/09/14 & 01:37 & 0.113 & 332 & & 1995/09/14 & 01:21 & 0.113 & 264 & & -  & -  \\
"         & 1995/09/14 & 22:53 & 0.149 & 321 & & 1995/09/14 & 23:09 & 0.149 & 339 & & -  & -  \\
"         & 1995/09/15 & 01:26 & 0.153 & 329 & & 1995/09/15 & 01:21 & 0.153 & 358 & & -  & -  \\
\noalign{\smallskip}
AR Lac    & 1995/09/12 & 22:16 & 0.405 & 360 & & 1995/09/12 & 21:57 & 0.399 & 421 & & 94,95a & 94,96a \\
"         & 1995/09/13 & 23:39 & 0.939 & 346 & & 1995/09/13 & 23:23 & 0.933 & 415 & & -  & -  \\
"         & 1995/09/14 & 22:49 & 0.425 & 339 & & 1995/09/14 & 22:42 & 0.423 & 407 & & -  & -  \\
\noalign{\smallskip}
KZ And    & 1995/09/14 & 23:29 & 0.145 & 336 & & 1995/09/14 & 23:51 & 0.150 & 367 & & -  & 94,95c,96a \\
\noalign{\smallskip}
KT Peg    & 1995/09/13 & 00:07 & 0.693 & 221 & & 1995/09/13 & 00:13 & 0.694 & 271 & & -  & 95c,96a \\
"         & 1995/09/15 & 01:19 & 0.024 & 313 & & 1995/09/15 & 01:13 & 0.023 & 368 & & -  & -  \\
\noalign{\smallskip}
II Peg    & 1995/09/12 & 23.02 & 0.575 & 300 & & 1995/09/12 & 23:23 & 0.577 & 298 & & -  & -  \\ 
"         & 1995/09/13 & 00:58 & 0.587 & 138 & & 1995/09/13 & 01:18 & 0.589 & 240 & & -  & -  \\
"         & 1995/09/14 & 00:57 & 0.735 & 265 & & 1995/09/14 & 01:14 & 0.737 & 324 & & -  & -  \\
"         & 1995/09/14 & 03:07 & 0.749 & 315 & & 1995/09/14 & 02:51 & 0.747 & 328 & & -  & -  \\
"         & 1995/09/14 & 05:00 & 0.760 & 266 & & 1995/09/14 & 05:16 & 0.762 & 263 & & -  & -  \\
"         & 1995/09/14 & 23:22 & 0.874 & 318 & & 1995/09/14 & 23:15 & 0.873 & 324 & & -  & -  \\ 
"         & 1995/09/15 & 01:56 & 0.890 & 332 & & 1995/09/15 & 01:49 & 0.889 & 352 & & -  & -  \\ 
"         & 1995/09/15 & 04:38 & 0.907 & 298 & & 1995/09/15 & 04:31 & 0.906 & 311 & & -  & -  \\ 
\noalign{\smallskip}
\hline
\end{tabular}

\vspace{0.5cm}
{\scriptsize
94:  Fern\'andez-Figueroa et al. (1994),
95a: Montes et al. (1995a),
95b: Montes et al. (1995b),
95c: Montes et al. (1995c),
96a: Montes et al. (1996a),
}

\end{flushleft}
\end{table*}

\begin{table*}
\caption[]{Stellar parameters
\label{tab:par}}
\begin{flushleft}
\scriptsize
\begin{tabular}{l l c c c c c l l c }
\noalign{\smallskip}
\hline
\noalign{\smallskip}
{HD} & {Name} & {T$_{\rm sp}$} & {SB} & {R} &
 {d} & {V-R} & {P$_{\rm orb}$} & {P$_{\rm rot}$} & Vsin{\it i}\\
             &    &     &    & (R$_\odot$) & (pc) &   & (days) &
 (days) & (km s$^{-1}$)  \\
\noalign{\smallskip}
\hline
\noalign{\smallskip}
1833   & BD Cet  &  K1III        & 1  & $\geq$10       & 71  &  0.81
& 35.1   & 34.46 & 15 \\
7672   & AY Cet  &   WD/G5III    & 1  & 0.012/15  & 66.7  &  0.69   &  56.824
& 77.22 & 4 \\
8357   & AR Psc &   G7V/K1IV$^1$ & 2  & /$\geq$1.5 & 17  &   0.74$^1$  &
14.30226$^1$ & 12.245 & 6.5/3.5$^1$  \\
12545  & XX Tri  & K0III       & 1  & $\geq$8  & 310 &  [0.62]  &  23.9824
& 24.3 & 17  \\
 21242 & UX Ari  &  G5V/K0IV     & 2  & 0.93/$\geq$4.7 & 50 &  0.70/0.54
& 6.43791 & $\approx$P$_{\rm orb}$ & 6/37 \\
22468 & V711 Tau &  G5IV/K1IV    & 2  & 1.3/3.9 & 36 & 0.62/0.75  &
 2.83774 & 2.841 & 13/38 \\
283750 &V833 Tau &  dK5e         & 1  & $\geq$0.22   & 16.7 & 0.69
 & 1.7878  & 1.797 & 6.3  \\
37824  & V1149 Ori &  K1III      & 1  & $\geq$11     & [164] & 0.90  & 53.58
& 54.1 & 11 \\
341475  & MM Her   & G2/K0IV     & 2 & 1.58/2.83 & 190 & [/0.64] & 7.960322 
& 7.936  & 10/18  \\
166181  & V815 Her & G5V/[M1-2V] & 1 & 0.93:/     & 31 &   0.54  &  1.8098368
& 1.8 & 27/ \\
234677  &  BY Dra  & K4V/K7.5V   &  2 & 1.2-1.4/   & 15.6 & 1.10 &  
5.975112 & 3.827 & 8.0/7.4  \\
175742  &  V775 Her &K0V/[K5-M2V]&  1 & 0.85/      & 24   &   0.80 &  
2.879395 & 2.898 & 15/ \\
178450  &  V478 Lyr &G8V/[dK-dM] &  1 & $\geq$0.9  & 26   &   0.65  &
 2.130514 & 2.185 & 21/ \\
209813  &  HK Lac   & F1V/K0III  & 1  & -          & 150  &  0.75   &
24.4284 & 24.4284 & /15  \\
210334  &  AR Lac   & G2IV/K0IV  & 2  & 1.8/3.1    & 47   &  0.77   &
1.98322195 & 1.98322195 & 46/81 \\
218738  &  KZ And   & dK2/dK2    &  2 & $\geq$0.74/ &[$\approx$23]
&  [0.74/0.74] & 3.032867 & 3.03 & 12.3/11.6   \\
%
%
222317  & KT Peg    & G5V/K6V    &  2 & 0.93/0.72  & 25 & [0.54/ ]
& 6.20199 & 6.092  & 8/5 \\
224085  &  II Peg   & K2-3V-IV   & 1  & 2.2        & 29.4 &  0.89   &
6.724183  & 6.718 & 21  \\
%
\noalign{\smallskip}
\hline
\noalign{\smallskip}
\end{tabular}

$^1$ Parameters from Fekel (1996)

\end{flushleft}
\end{table*}

\section{Observations and Data Reduction}

Observations in the H$\alpha$ and 
Na~{\sc i} D$_{1}$, D$_{2}$, He~{\sc i} D$_{3}$ line regions
have been obtained during
three nights (1995 September 13-15) with
the Isaac Newton Telescope (INT) at the
Observatorio del Roque de Los Muchachos (La Palma, Spain)  using the
Intermediate Dispersion Spectrograph (IDS) with grating H1800V, camera 500
and a 1024~x~1024 pixel TEK3 CCD as detector.
The reciprocal  dispersion  achieved  is
 0.24~\AA/pixel
which yields a spectral resolution of 0.48~\AA$\ $ and a useful wavelength
range of 250~\AA$\ $ centered at 6563\AA$\ $ (H$\alpha$)
and 5876\AA$\ $ (He{\sc i} D$_{3}$) respectively.

The spectra have been extracted using the standard
reduction procedures in the IRAF
%
%
 package (bias subtraction,
flat-field division, and optimal extraction of the spectra).
The wavelength calibration was obtained by taking
spectra of a Cu-Ar lamp.
Finally, the spectra have been normalized by 
a polynomial fit  to the observed continuum.

In Table~\ref{tab:obslog} we give the observing log. 
For each star we list
the date, UT, orbital phase ($\varphi$) and signal to noise ratio (S/N)
obtained for each observation in both spectral regions.
Where appropriate, we also give the reference 
of our previous observation of these 
systems in the H$\alpha$ and Ca~{\sc ii} H \& K lines.

In Table~\ref{tab:par} we show the HD number, name and the adopted
 stellar parameters (from CABS or the references given in the table) for the
 18 chromospherically active binary systems selected.

We have obtained the chromospheric contribution in
H$\alpha$ Na~{\sc i} D$_{1}$, D$_{2}$,  and He~{\sc i} D$_{3}$ lines using
the spectral subtraction technique described in
detail by Montes et al. (1995a,~c).

The synthesized spectra were
constructed using artificially rotationally broadened, radial-velocity
shifted, and weighted spectra of inactive stars
chosen to match the spectral types and luminosity classes
of both components of the active
system under consideration.
The reference stars used have been observed in this campaign and previous
observational seasons with similar spectral resolution 
(see the spectral library of Montes et al. 1997).

In some case, the difference spectrum obtained appears noisier than 
expected from the observation S/N ratio ($\approx$~300) due to
small differences in spectral type between active and reference star, or to
non appropriate evaluation of the rotational broadening 
and/or of the Doppler shift. 
In addition, in some spectra telluric lines also appear in the difference 
spectrum.
This noise in the the difference spectrum 
have been evaluated as the mean standard deviation ($\sigma$) 
in the regions outside the chromospheric features.
We have obtained values of $\sigma$ in the range 0.01-0.03 which
could be important in low active star but in the more active stars
the errors in the excess H$\alpha$ EW are small.
We have considered as a clear detection of excess emission or absorption
in H$\alpha$, Na~{\sc i} D$_{1}$, D$_{2}$, and He~{\sc i} D$_{3}$ only
when these features in the difference spectrum are 
larger than 3~$\sigma$. 

Table~\ref{tab:measures} gives the  H$\alpha$ line parameters, measured in the
observed and subtracted spectra of the sample. 
Column (2) of this Table gives the orbital phase ($\varphi$)
for each spectrum,
and  in column (3), H and C mean emission belonging to hot and
cool component respectively, and T means that at these phases
the spectral features cannot be deblended.
Column (4) gives the contributions for the hot and cool component
to the total continuum (S$_{\rm H}$ and S$_{\rm C}$).
Column (5) describes the observed H$\alpha$ profile, i.e. if the
line is in absorption (A) in emission (E) or totally filled by
emission (F).
Columns (6), (7), (8) give the following parameters 
measured in the observed spectrum:
the full width at half maximum (W$_{\rm obs}$);
the residual intensity, R$_{{\rm c}}$; and the H$\alpha$ core flux, F(1.7\AA), 
measured as the residual area below the central 1.7~\AA$\ $ passband.
The last four Columns give he following parameters
measured in the subtracted spectrum:
the full width at half maximum
(W$_{\rm sub}$), the peak emission intensity (I),
 the excess H$\alpha$ emission equivalent width (EW( H$\alpha$)), and 
absolute fluxes at the stellar surface logF$_{\rm S}$(H$\alpha$) obtained
with the calibration of Pasquini \& Pallavicini (1991) as a function of 
(V~-~R), very similar values of F$_{\rm S}$(H$\alpha$) are obtained
using the more recently calibration of Hall (1996)  as a function of
(V~-~R) and (B~-~V).
For a more detailed description of the parameters given  in this table
see our previous study of the excess H$\alpha$ emission in active binaries
(Montes et al. 1995a).

In Table~\ref{tab:measures_nb} we list the parameters (I, FWHM, EW) 
of the broad and narrow components used in the two Gaussian components fit 
to the H$\alpha$ subtracted emission profile, which we have performed in the 
stars that present broad wings. See the comments for each individual star in 
Sect.~3 and the interpretation of these components given in Sect.~4.

\begin{figure*}
{\psfig{figure=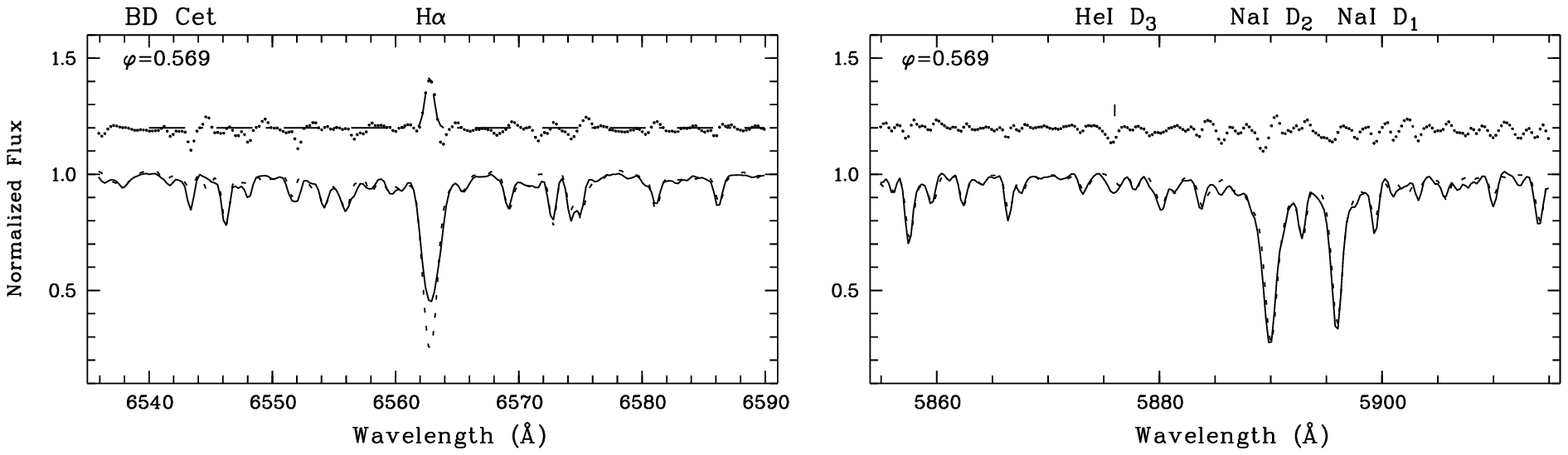,bbllx=288pt,bblly=73pt,bburx=450pt,bbury=627pt,height=5.6cm,width=18.0cm,angle=270,clip=}}
\caption[ ]{H$\alpha$, Na~{\sc i} D$_{1}$, D$_{2}$, and He~{\sc i} D$_{3}$ 
spectra of BD Cet
\label{fig:plot_bdcet_hahe} }
\end{figure*}

\begin{figure*}
{\psfig{figure=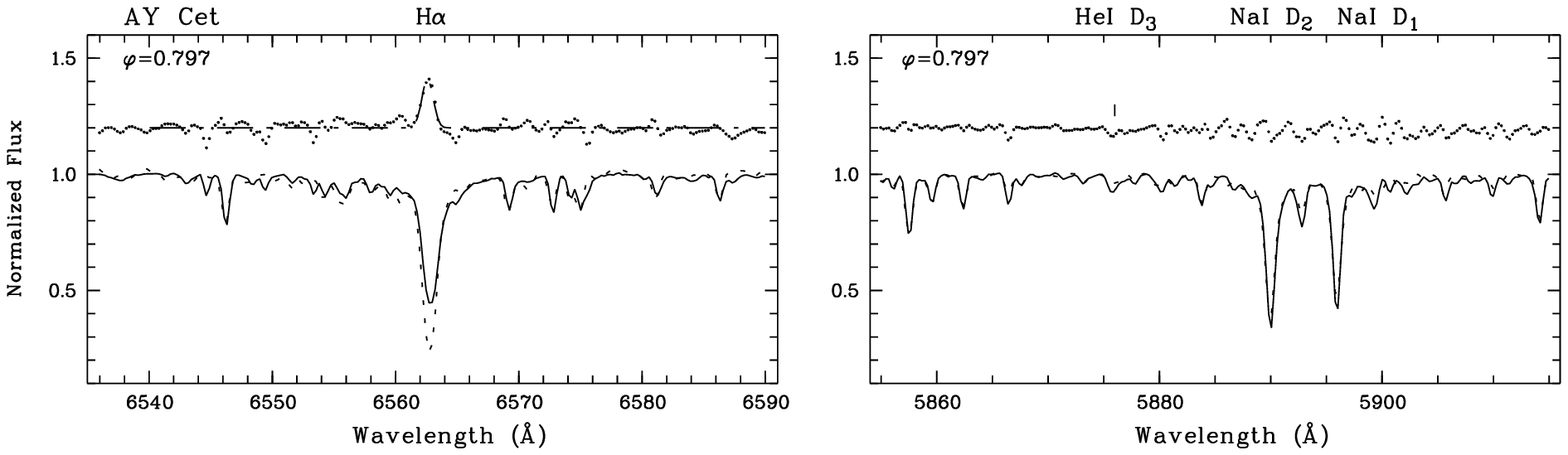,bbllx=288pt,bblly=73pt,bburx=450pt,bbury=627pt,height=5.6cm,width=18.0cm,angle=270,clip=}}
\caption[ ]{H$\alpha$, Na~{\sc i} D$_{1}$, D$_{2}$, and He~{\sc i} D$_{3}$ 
spectra of AY Cet
\label{fig:plot_aycet_hahe} }
\end{figure*}

\section{Individual results}

In the following we describe the H$\alpha$, Na~{\sc i} D$_{1}$, D$_{2}$,
and He~{\sc i} D$_{3}$ spectra of the stars of this sample.
The line profiles of each
chromospherically active binary system
are displayed in Fig.~1 to 19
 The name of the star, the orbital phase, and the expected
positions of the features for the hot (H) and cool (C)
 components are given in each figure.
For each system we plot the observed spectrum (solid-line), the
synthesized spectrum (dashed-line), the subtracted spectrum, additively
offset for better display (dotted line)
 and the Gaussian fit to the subtracted spectrum (dotted-dashed line).

\subsection{BD Cet (HD 1833)}

Single-lined spectroscopic binary classified as K1III + F by
Bidelman \&  MacConnell (1973). 
It presents strong Ca~{\sc ii} H \& K emission lines  centered at the 
absorption line (Montes et al. 1995c)
and the H$\alpha$ line as moderate absorption (Fekel  et al. 1986).

We have obtained one spectrum of this system 
at the orbital phase 0.569 (see Fig.~\ref{fig:plot_bdcet_hahe}).
The H$\alpha$ subtracted spectrum shows a weak excess emission.
In the Na~{\sc i} line region no detectable filling-in of the
D$_{1}$ and D$_{2}$ lines is present.
A clear absorption in the He~{\sc i} D$_{3}$ line 
appears in the subtracted spectrum.
In both spectral regions the spectrum is matched 
using a K2III as reference star.

\subsection{AY Cet (39 Cet, HD 7672, HR 373)}

AY Cet is a single-lined binary composed of a spotted G5III primary and a white
dwarf secondary. It presents strong Ca~{\sc  ii} H \& K
emission lines (Montes et al. 1995c) and a filled in absorption H$\alpha$ line
(Fekel et al. 1986;  Strassmeier  et  al. 1990).

We present here one observation of this system at the orbital phase 0.797
(see Fig.~\ref{fig:plot_aycet_hahe}). 
In the H$\alpha$ line region we have used a G8III reference star to perform 
the spectral subtraction and the
subtracted spectrum obtained shows a weak excess emission.
No detectable filling-in of the Na~{\sc i} D$_{1}$ and D$_{2}$ lines
has been found, however, a weak absorption in the He~{\sc i} D$_{3}$ line 
appears in the subtracted spectrum.

\begin{table*}
\caption[]{H$\alpha$ line measures in the observed and
subtracted spectrum
\label{tab:measures}}
\begin{flushleft}
\scriptsize
\begin{tabular}{lcccccccccccc}
\noalign{\smallskip}
\hline
\noalign{\smallskip}
 &   &   &  & 
\multicolumn{4}{c}{Observed H$\alpha$ Spectrum} &\ &
\multicolumn{4}{c}{Subtracted H$\alpha$ Spectrum} \\
\cline{5-8}\cline{10-13}
\noalign{\smallskip}
 {Name} &  {$\varphi$} & {E} &  S$_{\rm H}$/S$_{\rm C}$  & H$\alpha$ &
 {W$_{\rm obs}$} & {R$_{\rm c}$ } & {F(1.7\AA)} &  &
 {W$_{\rm sub}$} & {I} & {EW} & {$\log {\rm F}_{\rm S}$} \\
  & &  &   &   & {\scriptsize (\AA) } &  &  & &
 {\scriptsize (\AA)} &  & {\scriptsize  (\AA)} &  \\
\noalign{\smallskip}
\hline
\noalign{\smallskip}
%
 BD Cet   & 0.569 & - &   -       & A & 1.80 & 0.452 & 0.928 & &
0.82 & 0.221 & 0.193 & 5.75 \\ 
 AY Cet   & 0.797 & - &   -       & A & 1.62 & 0.446 & 0.940 & &
1.08 & 0.214 & 0.245 & 6.02 \\ 
 AR Psc   & 0.373 & - & 0.20/0.80 & E & -    & 1.065 & 1.715 & &
 1.47 & 0.599 & 1.071 & 6.58 \\
 "        & 0.443 & - & 0.20/0.80 & E & -    & 1.063 & 1.680 & &
 1.61 & 0.653 & 1.403 & 6.70 \\
 "        & 0.519 & - & 0.20/0.80 & E & -    & 1.059 & 1.736 & &
 1.46 & 0.676 & 1.211 & 6.64 \\ 
 "        & 0.524 & - & 0.20/0.80 & E & -    & 1.070 & 1.755 & &
 1.47 & 0.687 & 1.263 & 6.66 \\ 
 XX Tri   & 0.401 & - &  -        & E & 2.68 & 1.387 & 2.301 & &
1.80  & 1.089 & 2.599 & 6.92 \\
 UX Ari   & 0.419 & C & 0.30/0.70 & E & 1.26 & 1.225 & 1.966 & &
1.89 & 0.722 & 1.798 & 6.85 \\
 "        & 0.438 & C & 0.30/0.70 & E & 1.07 & 1.101 & 1.799 & &
1.82 & 0.626 & 1.405 & 6.74 \\
 "        & 0.576 & C & 0.30/0.70 & E & 1.41 & 1.159 & 1.818 & &
1.83 & 0.609 & 1.476 & 6.76 \\
 "        & 0.736 & C & 0.30/0.70 & E & 3.17 & 1.533 & 2.504 & &
2.51 & 0.996 & 3.286 & 7.11 \\
 V711 Tau & 0.922 & C & 0.16/0.84 & E & 3.36 & 1.315 & 2.084 & &
 2.48 & 0.837 & 2.710 & 6.96 \\
 "        & 0.261 & C & 0.16/0.84 & E & 3.63 & 1.264 & 2.095 & &
 2.26 & 0.759 & 2.524 & 6.93 \\ 
 "        & 0.280 & C & 0.16/0.84 & E & 3.51 & 1.255 & 2.081 & &
 2.24 & 0.763 & 2.421 & 6.92 \\
 "        & 0.606 & C & 0.16/0.84 & E & 2.93 & 1.342 & 2.185 & &
 2.20 & 0.842 & 2.641 & 6.95 \\ 
 "        & 0.641 & C & 0.16/0.84 & E & 2.88 & 1.328 & 2.164 & &
 2.24 & 0.812 & 2.382 & 6.91 \\
 V833 Tau & 0.762 & - & -         & E & 1.61 & 1.087 & 1.808 & & 
1.37 & 0.754 & 1.101 & 6.37 \\
 "        & 0.319 & - & -         & E & -    & 1.078 & 1.738 & &
1.35 & 0.709 & 1.018 & 6.34 \\
 "        & 0.851 & - & -         & E & 1.56 & 1.085 & 1.790 & &
1.35 & 0.745 & 1.070 & 6.36 \\
 V1149 Ori& 0.439 & - &   -       & A & 2.00 & 0.476 & 0.953 & &
0.91 & 0.269 & 0.259 & 5.75 \\
 MM Her   & 0.498 & C & 0.59/0.41 & A & 1.65 & 0.561 & 1.118 & &
 1.42 & 0.320 & 0.590 & 6.38 \\          
 "        & 0.630 & H & 0.50      & A & -    & 0.588 & 1.101 & &
 -    & 0.000 & 0.000 & 0.00 \\    
          &       & C & 0.50      & A & -    & 0.846 & 1.490 & &
 1.46 & 0.302 & 0.541 & 6.34 \\      
 "        & 0.745 & H & 0.50      & A & -    & 0.603 & 1.134 & &
 -    & 0.000 & 0.000 & 0.00 \\      
          &       & C & 0.50      & A & -    & 0.831 & 1.476 & &
 1.41 & 0.270 & 0.494 & 6.30 \\ 
 V815 Her & 0.978 & H &   -       & F & -    & 0.878 & 1.508 & &
 1.75 & 0.535 & 1.279 & 6.95 \\
  "       & 0.520 & H &   -       & A & -    & 0.739 & 1.314 & &
 1.65 & 0.396 & 0.875 & 6.79 \\
  "       & 0.099 & H &   -       & A & -    & 0.787 & 1.390 & &
 1.80 & 0.445 & 1.095 & 6.89 \\
 BY Dra   & 0.684 & T & 0.70/0.30 & E & 1.39 & 1.202 & 1.919 & &
      &       & 1.357 &      \\  
          &       & H & 0.70      & E &      &       &       & &
1.36 & 0.540 & 0.780 & 6.20 \\
          &       & C & 0.30      & E &      &       &       & &
1.70 & 0.319 & 0.577 & 5.73 \\
 "        & 0.839 & T & 0.70/0.30 & E & 1.70 & 1.270 & 2.039 & &
      &       & 1.455 &      \\  
          &       & H & 0.70      & E &      &       &       & &
1.38 & 0.548 & 0.803 & 6.22 \\
          &       & C & 0.30      & E &      &       &       & &
1.89 & 0.324 & 0.652 & 5.78 \\
 V775 Her & 0.394 & H &   -       & E & -    & 1.044 & 1.719 & &
1.39 & 0.745 & 1.103 & 6.51 \\
          &       & C &   -       & E & -    & 1.027 & 1.690 & &
1.02 & 0.138 & 0.149 & -    \\
 V478 Lyr & 0.953 & H &   -       & A & 3.36 & 0.763 & 1.353 & & 
1.53 & 0.446 & 0.727 & 6.55 \\
 HK Lac   & 0.067 & C &   -       & A & 1.66 & 0.611 & 1.178 & &
 1.31 & 0.366 & 0.765 & 6.43 \\ 
 "        & 0.070 & C &   -       & A & 1.75 & 0.610 & 1.170 & & 
      &       &       &      \\
 "        & 0.110 & C &   -       & A & 1.61 & 0.657 & 1.243 & & 
 1.32 & 0.411 & 0.887 & 6.49 \\
 "        & 0.113 & C &   -       & A & 1.61 & 0.647 & 1.237 & &
      &       &       &      \\   
 "        & 0.149 & C &   -       & A & 1.61 & 0.664 & 1.257 & &
 1.37 & 0.420 & 0.929 & 6.51 \\ 
 "        & 0.153 & C &   -       & A & 1.57 & 0.661 & 1.244 & &
      &       &       &      \\
%
 AR Lac   & 0.405 & H & 0.36      & A & -    & 0.787 & 1.401 & & 
 1.92 & 0.083 & 0.169 & 5.98 \\
          &       & C & 0.64      & A & -    & 0.774 & 1.335 & &
2.037 & 0.114 & 0.248 & 6.00 \\
 "        & 0.425 & H & 0.37      & A & -    & 0.696 & 1.241 & & 
 1.44 & 0.062 & 0.095 & 5.73 \\
          &       & C & 0.63      & A & -    & 0.675 & 1.167 & &
 0.92 & 0.018 & 0.018 & 4.86 \\
 "        & 0.939 & H & 0.27      & A & -    & -     & -     & & 
 -    & -     & -     & -    \\
          &       & C & 0.73      & A & -    & -     & -     & &
 -    & 0.050 & 0.104 & 5.59 \\
 KZ And   & 0.145 & 1 & 0.58      & A & 1.81 & 0.780 & 1.404 & &
1.06 & 0.209 & 0.237 & 5.94 \\
          &       & 2 & 0.42      & A & -    & 0.884 & 1.517 & &
1.22 & 0.207 & 0.269 & 5.99 \\
 KT Peg   & 0.693 & C & 0.95/0.05 & A & 1.85 & 0.349 & 0.786 & &
 -    & 0.000 & 0.000 & 0.00 \\ 
 "        & 0.024 & C & 0.95/0.05 & A & 1.77 & 0.354 & 0.792 & &
 -    & 0.000 & 0.000 & 0.00 \\ 
 II Peg   & 0.575 & - &   -       & E & 1.69 & 1.456 & 2.316 & & 
 1.62 & 1.134 & 2.127 & 6.66 \\
 "        & 0.587 & - &   -       & E & 1.69 & 1.456 & 2.302 & & 
 1.62 & 1.135 & 2.146 & 6.66 \\ 
 "        & 0.735 & - &   -       & E & 2.04 & 1.407 & 2.311 & &
 1.69 & 1.102 & 2.362 & 6.70 \\  
 "        & 0.749 & - &   -       & E & 2.02 & 1.393 & 2.295 & & 
 1.66 & 1.087 & 2.486 & 6.72 \\   
 "        & 0.760 & - &   -       & E & 2.63 & 1.505 & 2.484 & &
 1.80 & 1.198 & 3.692 & 6.88 \\
 "        & 0.874 & - &   -       & E & 1.87 & 1.279 & 2.102 & &
 1.57 & 0.956 & 1.963 & 6.62 \\    
 "        & 0.890 & - &   -       & E & 1.91 & 1.267 & 2.084 & &
 1.57 & 0.942 & 1.972 & 6.62 \\   
 "        & 0.907 & - &   -       & E & 1.85  & 1.288 & 2.096 & & 
 1.66 & 0.959 & 1.961 & 6.62 \\  
\noalign{\smallskip}
\hline
\noalign{\smallskip}
\end{tabular}
\end{flushleft}
\end{table*}

\subsection{AR Psc (HD 8357) }

This extremely active RS Cvn system is a
double-lined spectroscopic binary consisting of a K1IV primary and a G7V 
secondary (Fekel 1996).
In our previous observations of this system we have found
strong Ca~{\sc  ii} H \& K and H$\epsilon$ emissions 
from the cool component (FFMCC),
the H$\alpha$ line of the active component in emission above
the continuum (FFMCC and Montes et al. 1995a, b) and an important filling-in 
by chromospheric emission in the H$\beta$ line (Montes et al. 1995d).

\begin{figure*}
{\psscalefirst
\vspace{-0.5cm}
\hspace{-1.8cm}
{\psfig{figure=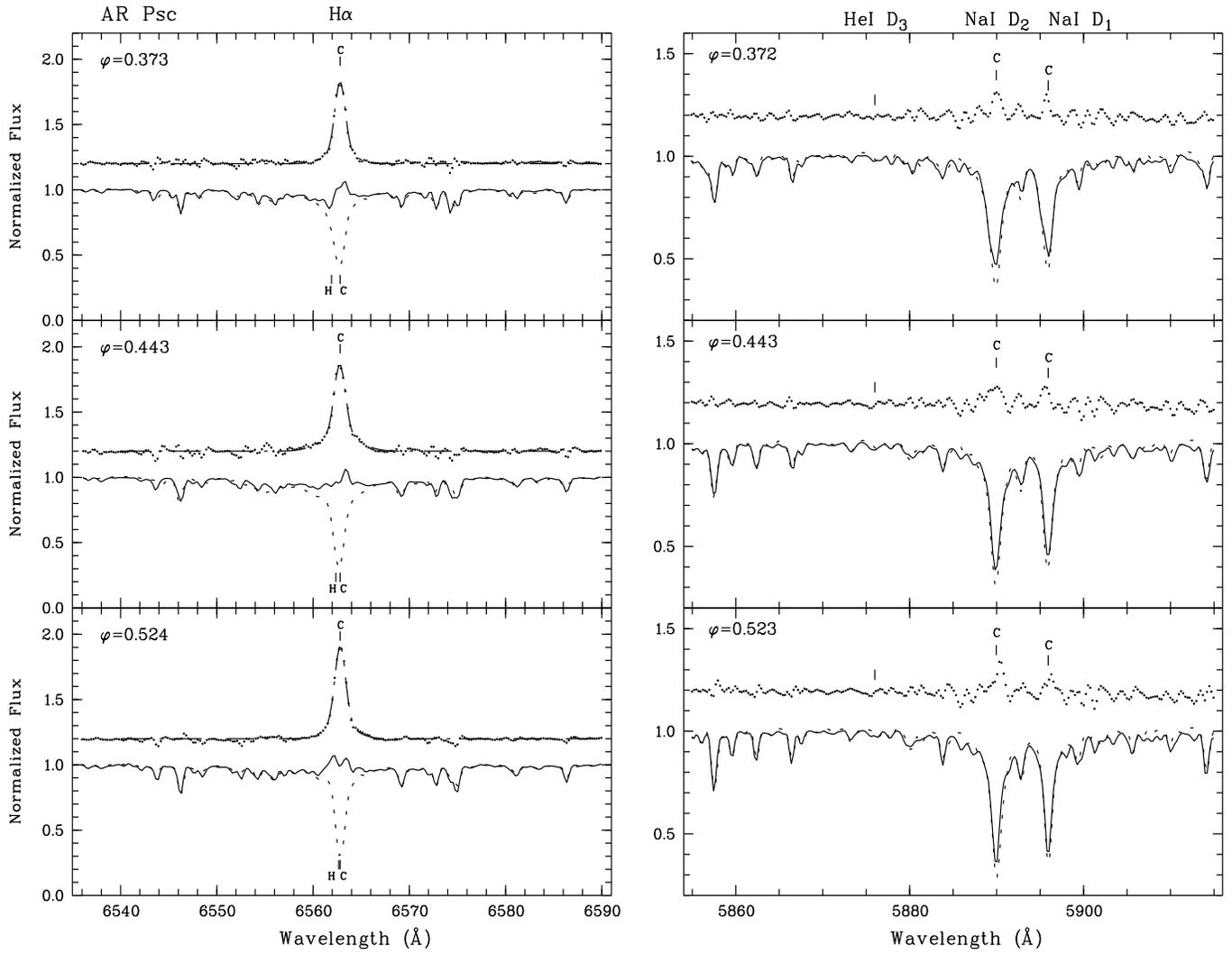,height=27.5cm,width=18.0cm,rheight=14.4cm,rwidth=6cm,angle=270}}
}
\caption[ ]{H$\alpha$, Na~{\sc i} D$_{1}$, D$_{2}$, and He~{\sc i} D$_{3}$ 
spectra of AR Psc
\label{fig:plot_arpsc_hahe} }
\end{figure*}

Now we have analysed
four spectra of this system in both spectral regions
at the orbital phases from 0.37 to 0.52 (see Fig.~\ref{fig:plot_arpsc_hahe}).
 The H$\alpha$ line of the active component appears in emission above
 the continuum, with a profile that changes with
the orbital phase owing to the different amounts of overlapping with 
the absorption of the other component.
By subtracting the synthesized spectrum, constructed with G6IV and K0IV
reference stars and a relative contribution of 0.2/0.8
we have found a large excess H$\alpha$ emission which is well matched  
using a two-components Gaussian fit.
In the Na~{\sc i} D$_{1}$ and D$_{2}$ lines an important excess 
emission is present in the subtracted spectra, 
however, the He~{\sc i} D$_{3}$ line does not appear in absorption. 

\begin{figure*}
{\psfig{figure=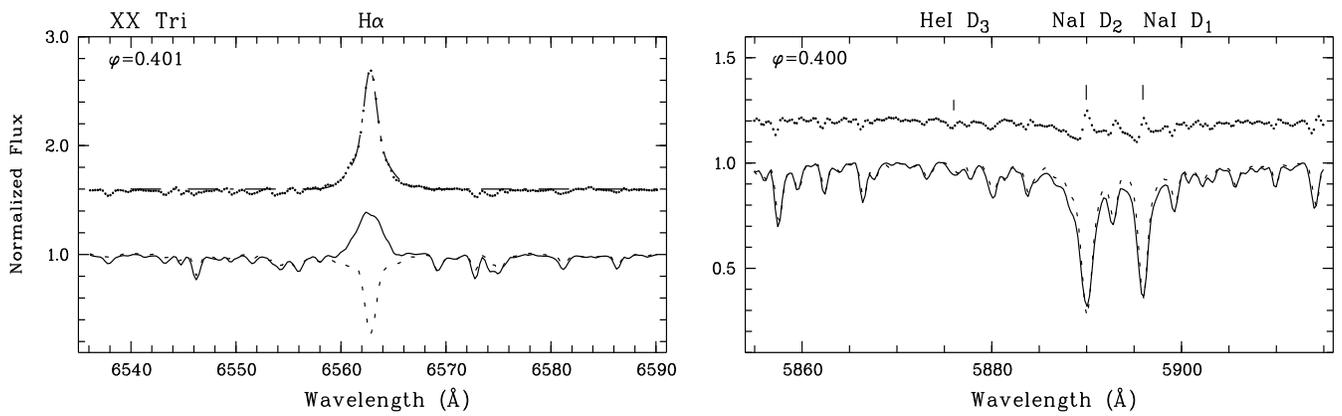,bbllx=288pt,bblly=73pt,bburx=450pt,bbury=627pt,height=5.6cm,width=18.0cm,angle=270,clip=}}
\caption[ ]{H$\alpha$, Na~{\sc i} D$_{1}$, D$_{2}$, and He~{\sc i} D$_{3}$ 
spectra of XX Tri (HD 12545)
\label{fig:plot_xxtri_hahe} }
\end{figure*}

\begin{figure*}
{\psscalefirst
\vspace{-0.5cm}
\hspace{-1.8cm}
{\psfig{figure=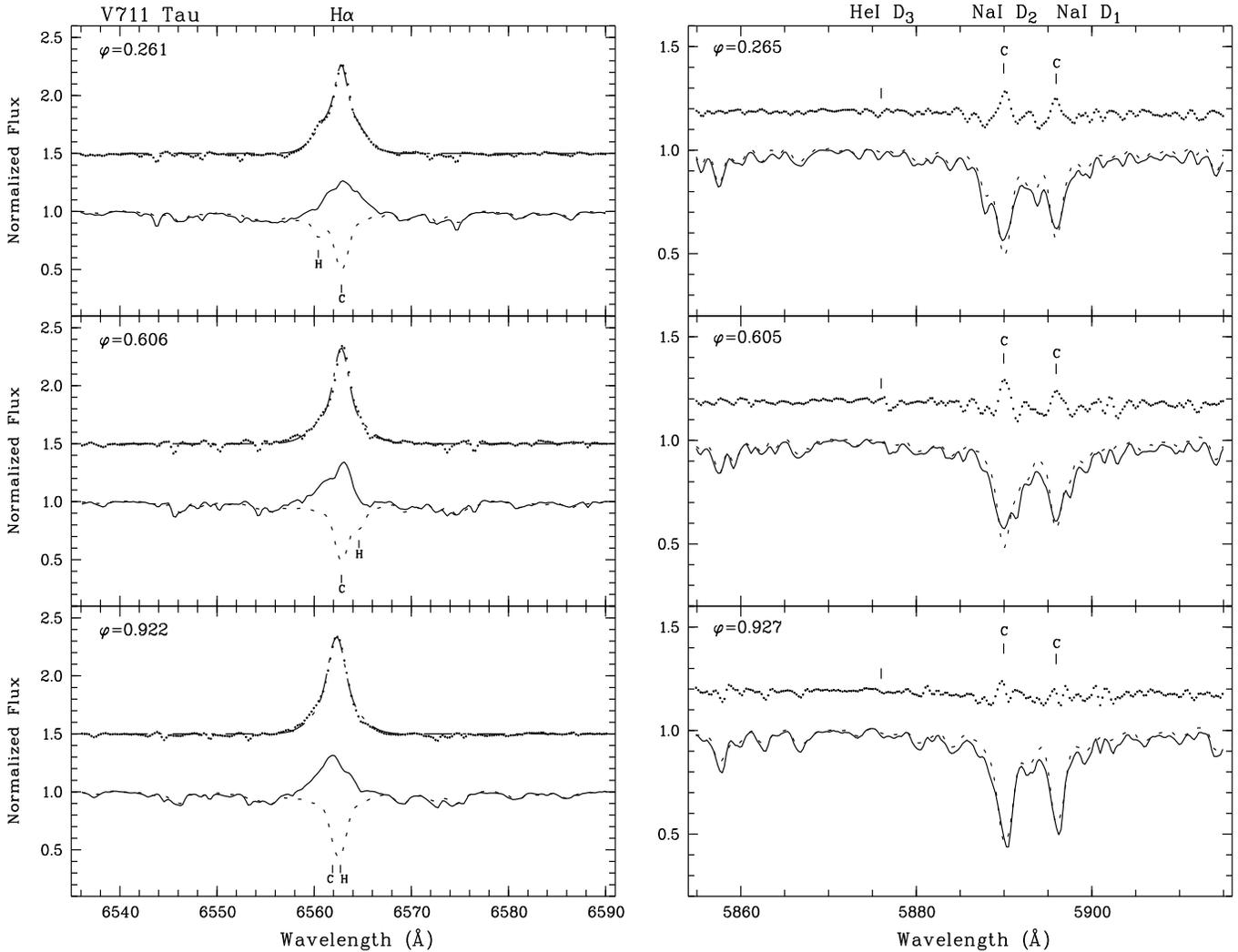,height=27.5cm,width=18.0cm,rheight=14.4cm,rwidth=6cm,angle=270}}
}
\caption[ ]{H$\alpha$, Na~{\sc i} D$_{1}$, D$_{2}$, and He~{\sc i} D$_{3}$ 
spectra of V711 Tau
\label{fig:plot_v711tau_hahe} }
\end{figure*}
\begin{figure*}
{\psscalefirst
\vspace{-0.5cm}
\hspace{-1.8cm}
{\psfig{figure=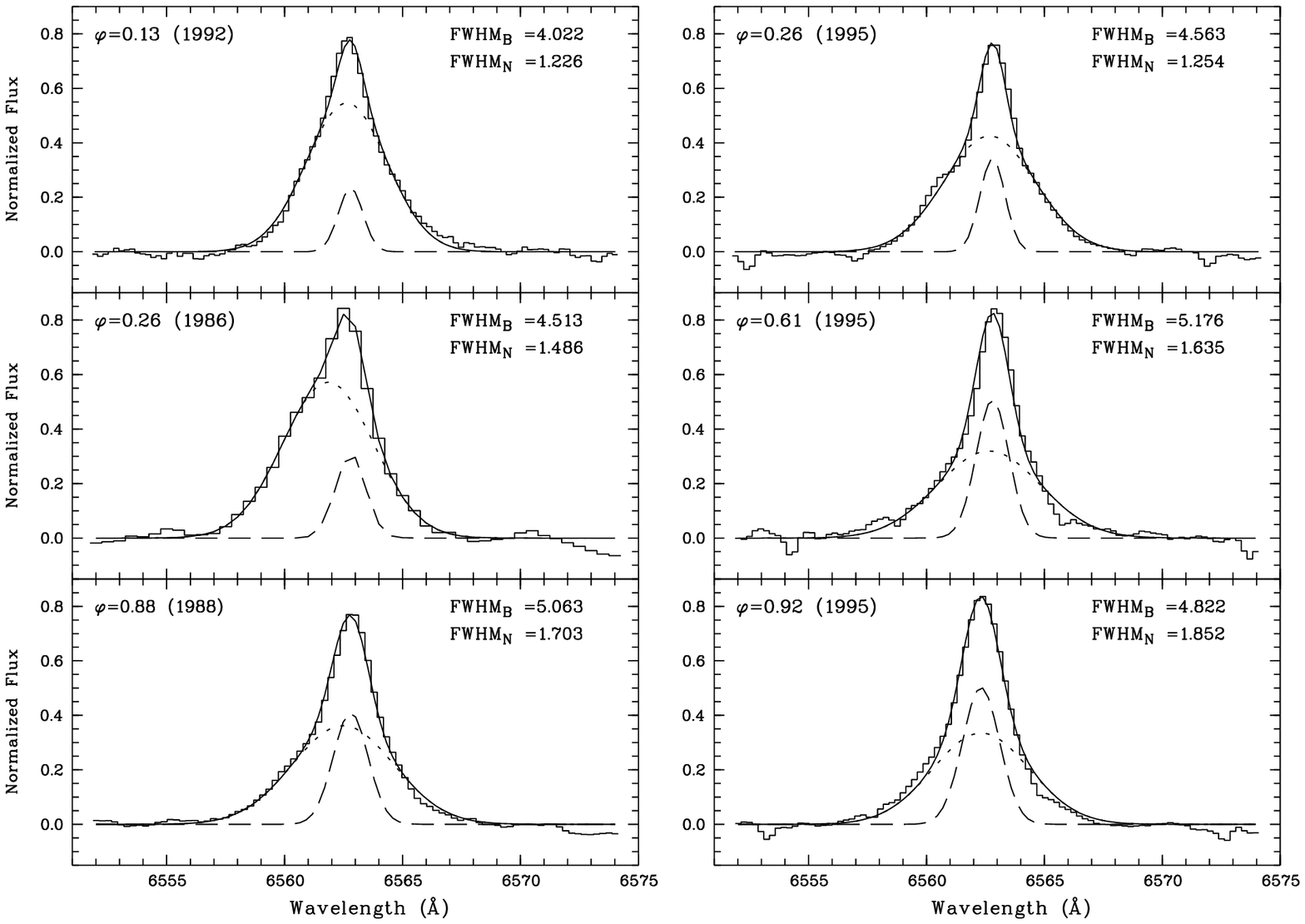,height=27.5cm,width=18.0cm,rheight=14.4cm,rwidth=6cm,angle=270}}
}
\caption[ ]{Subtracted H$\alpha$ profiles  of V711 Tau 
at different epochs and orbital phases (line in histogram from).
We have superposed the two Gaussian components fit (solid-line).
The sort-dashed-line represents the broad component 
and the large-dashed-line the narrow one.
\label{fig:plot_v711tau_hahe_nb} }
\end{figure*}

\begin{figure*}
{\psscalefirst
\vspace{-0.5cm}
\hspace{-1.8cm}
{\psfig{figure=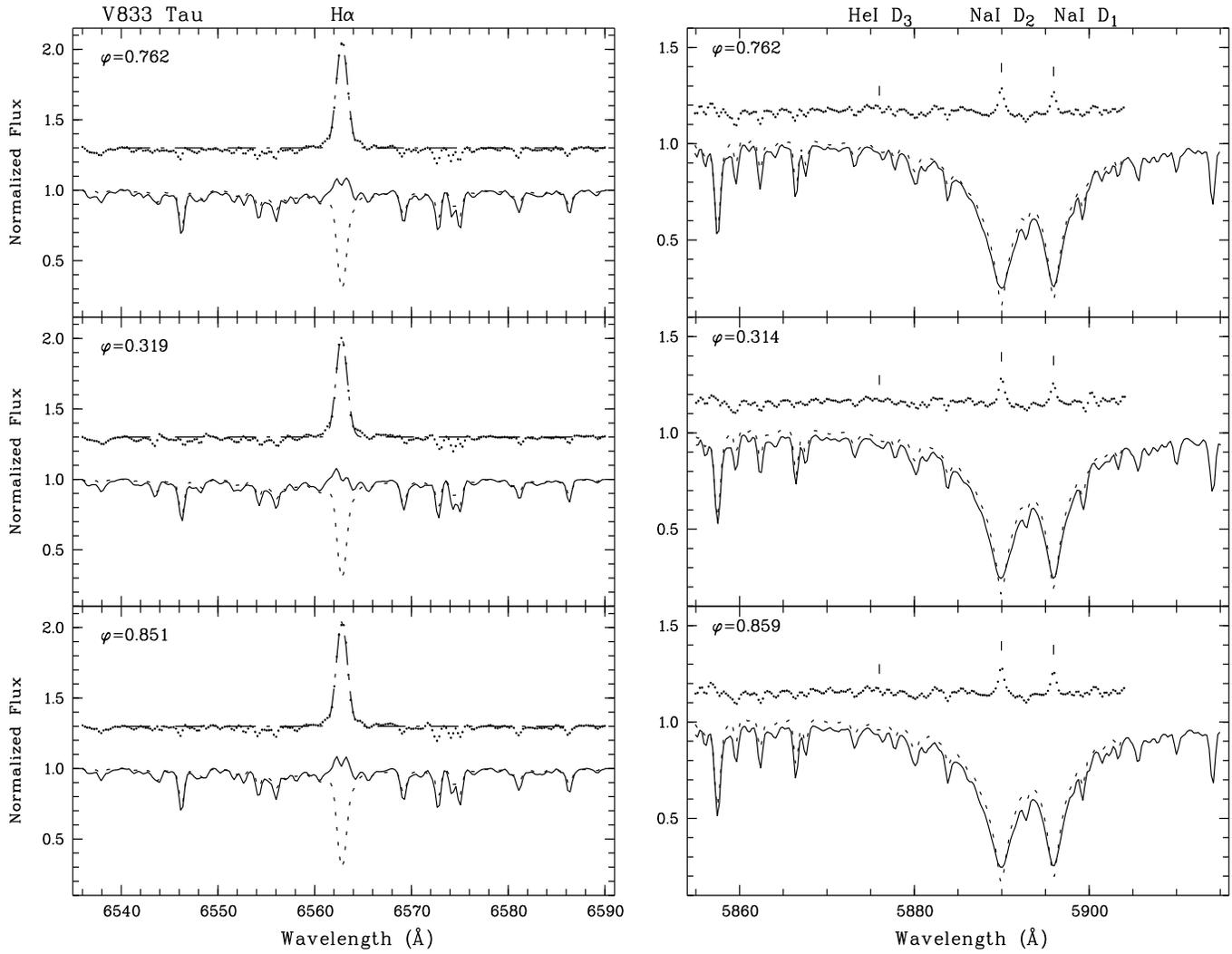,height=27.5cm,width=18.0cm,rheight=14.4cm,rwidth=6cm,angle=270}}
}
\caption[ ]{H$\alpha$, Na~{\sc i} D$_{1}$, D$_{2}$, and He~{\sc i} D$_{3}$ 
spectra of V833 Tau
\label{fig:plot_v833tau_hahe} }
\end{figure*}

\begin{figure*}
{\psfig{figure=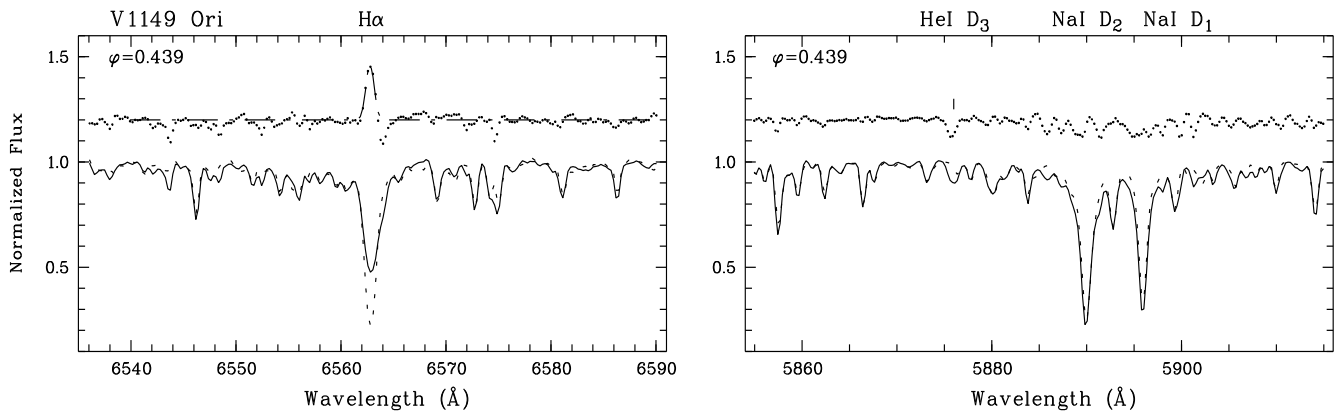,bbllx=288pt,bblly=73pt,bburx=450pt,bbury=627pt,height=5.6cm,width=18.0cm,angle=270,clip=}}
\caption[ ]{H$\alpha$, Na~{\sc i} D$_{1}$, D$_{2}$, and He~{\sc i} D$_{3}$ 
spectra of V1149 Ori
\label{fig:plot_v1149ori_hahe} }
\end{figure*}

\begin{figure*}
\hbox{
{\psfig{figure=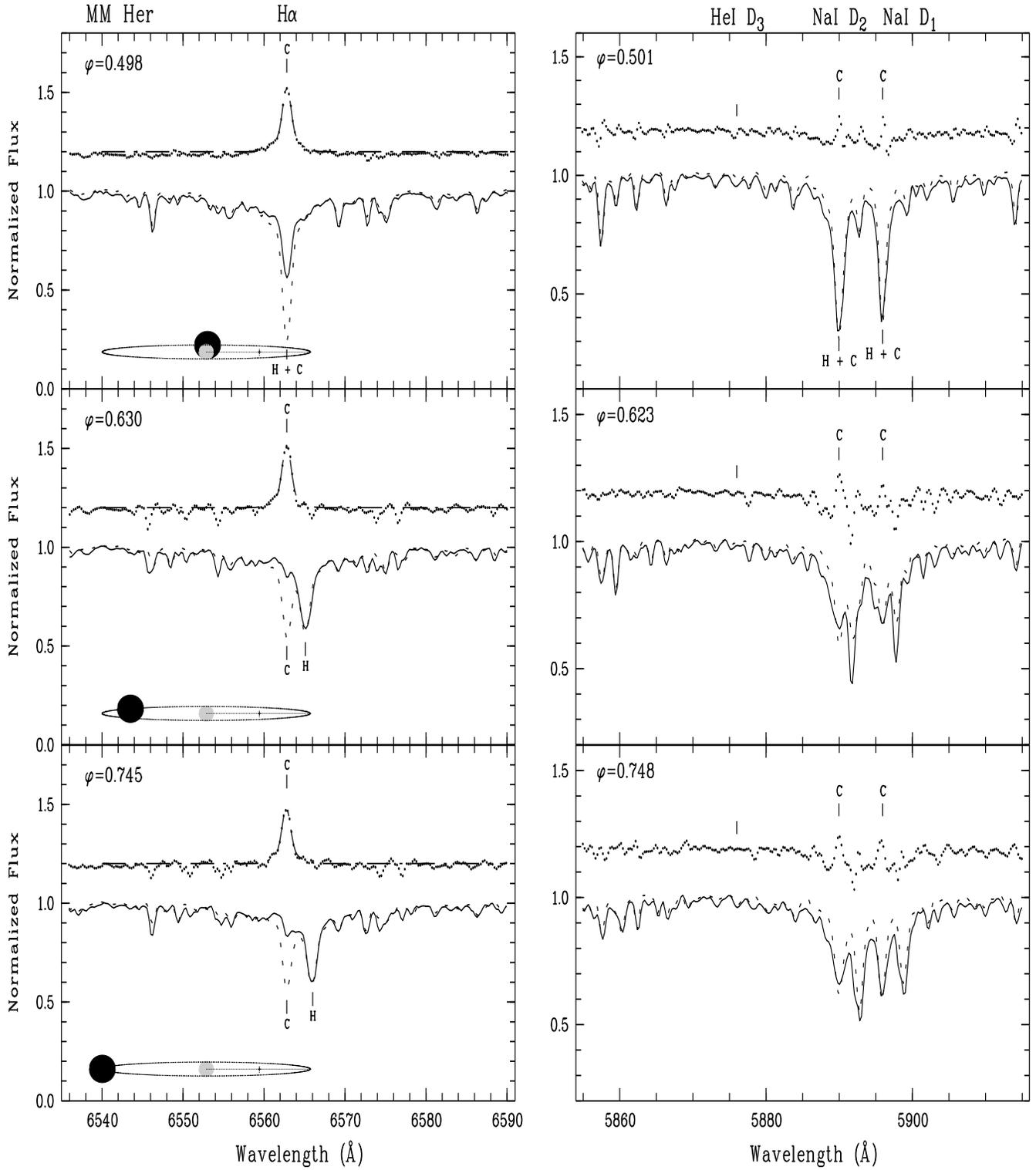,bbllx=44pt,bblly=73pt,bburx=450pt,bbury=627pt,height=20.5cm,width=18.0cm,rheight=20.5cm,angle=270,clip=}}
\hspace{-16.7cm}
\vbox{
{\psfig{figure=ds5809f9_a.ps,bbllx=70pt,bblly=85pt,bburx=344pt,bbury=202pt,height=1.8cm,width=4.0cm,clip=}}
\vspace{4.4cm}
{\psfig{figure=ds5809f9_b.ps,bbllx=70pt,bblly=85pt,bburx=344pt,bbury=202pt,height=1.8cm,width=4.0cm,clip=}}
\vspace{4.3cm}
{\psfig{figure=ds5809f9_c.ps,bbllx=70pt,bblly=85pt,bburx=344pt,bbury=202pt,height=1.8cm,width=4.0cm,clip=}}
\vspace{0.9cm}
}
}
\caption[ ]{H$\alpha$, Na~{\sc i} D$_{1}$, D$_{2}$, and He~{\sc i} D$_{3}$ 
spectra of MM Her
\label{fig:plot_mmher_hahe} }
\end{figure*}

\begin{figure*}
{\psscalefirst
\vspace{-0.5cm}
\hspace{-1.8cm}
{\psfig{figure=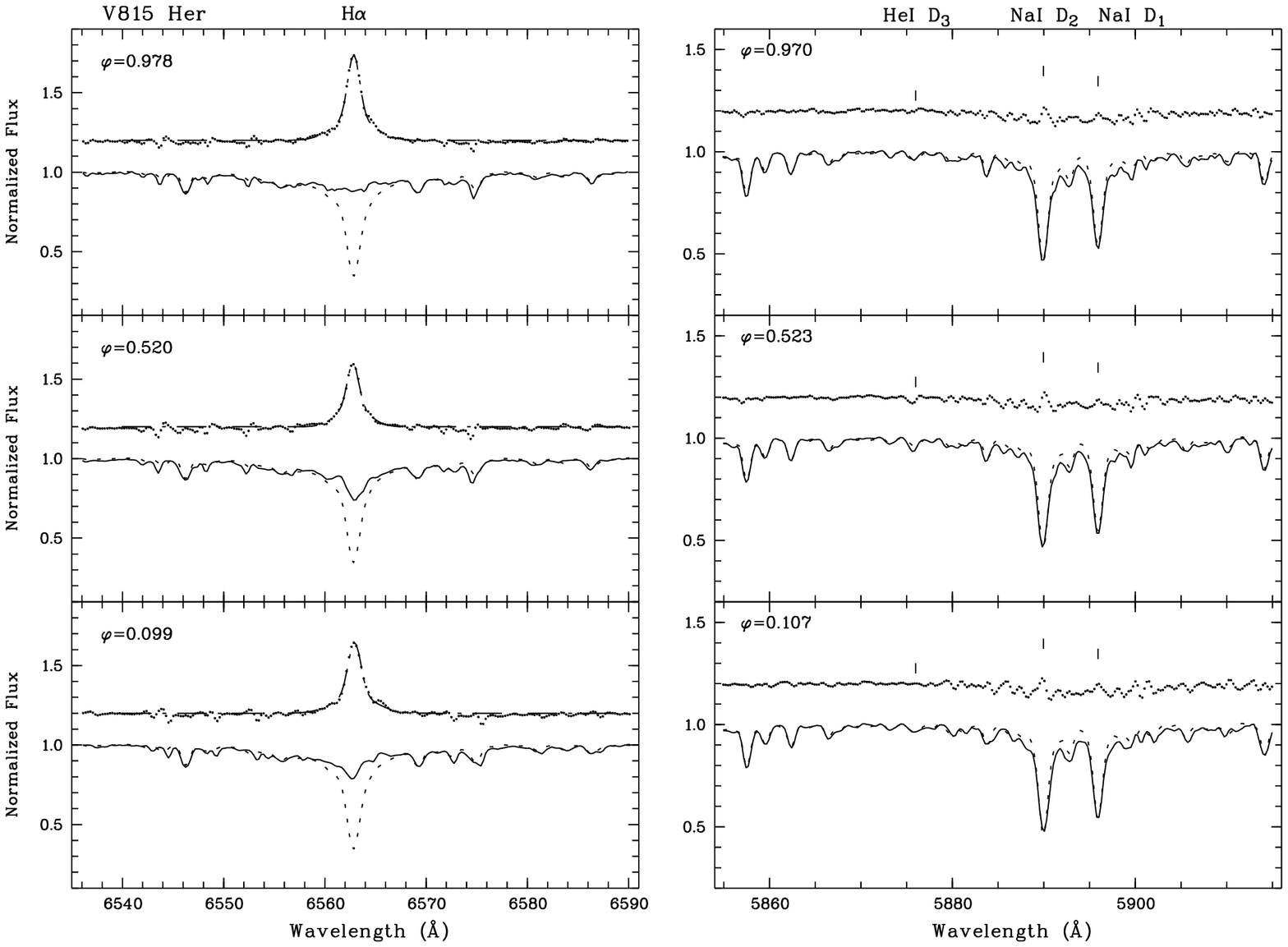,height=27.5cm,width=18.0cm,rheight=14.4cm,rwidth=6cm,angle=270}}
}
\caption[ ]{H$\alpha$, Na~{\sc i} D$_{1}$, D$_{2}$, and He~{\sc i} D$_{3}$ 
spectra of V815 Her
\label{fig:plot_v815her_hahe} }
\end{figure*}

\begin{figure*}
{\psfig{figure=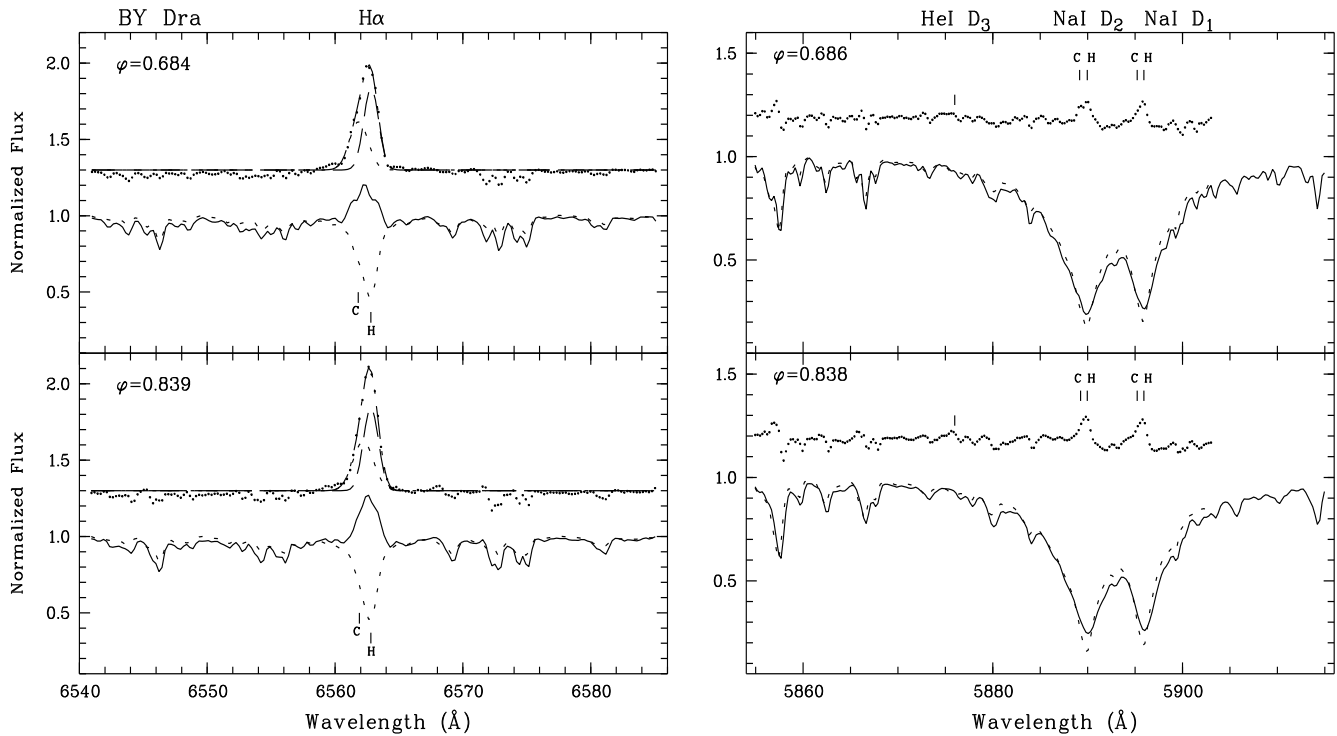,bbllx=44pt,bblly=73pt,bburx=329pt,bbury=627pt,height=10.0cm,width=18.0cm,angle=270,clip=}}
\caption[ ]{H$\alpha$, Na~{\sc i} D$_{1}$, D$_{2}$, and He~{\sc i} D$_{3}$ 
spectra of BY Dra. In the subtracted H$\alpha$ spectra we have superposed
the Gaussian fit used to deblend the contribution 
of the hot and cool components 
\label{fig:plot_bydra_hahe} }
\end{figure*}

\begin{figure*}
{\psfig{figure=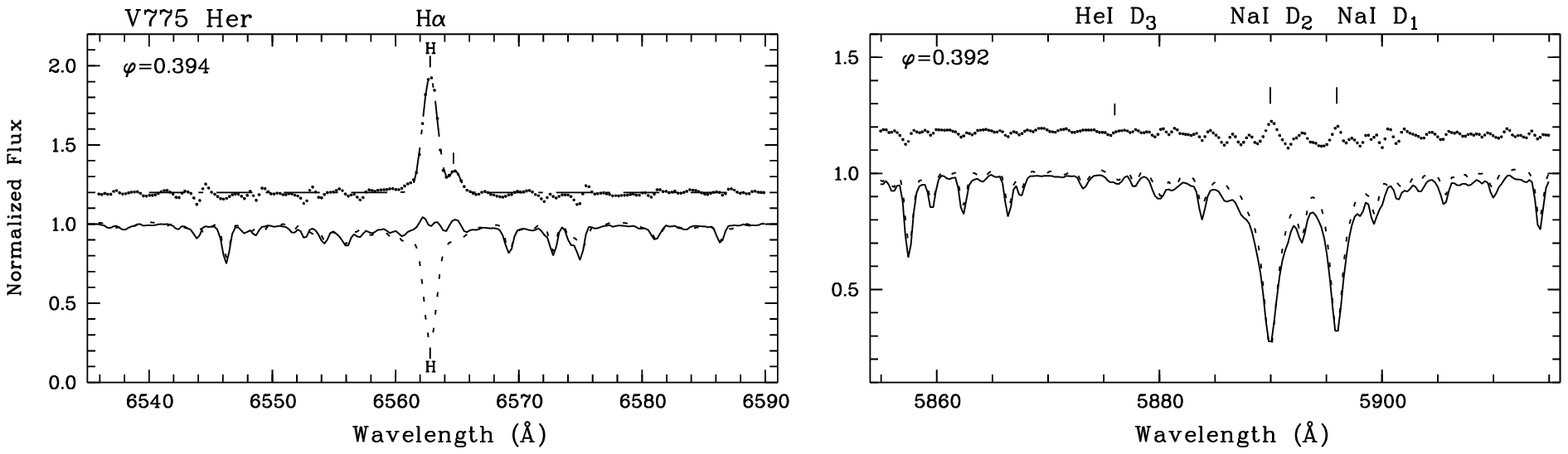,bbllx=288pt,bblly=73pt,bburx=450pt,bbury=627pt,height=5.6cm,width=18.0cm,angle=270,clip=}}
\caption[ ]{H$\alpha$, Na~{\sc i} D$_{1}$, D$_{2}$, and He~{\sc i} D$_{3}$ 
spectra of V775 Her
\label{fig:plot_v775her_hahe} }
\end{figure*}

\begin{figure*}
{\psfig{figure=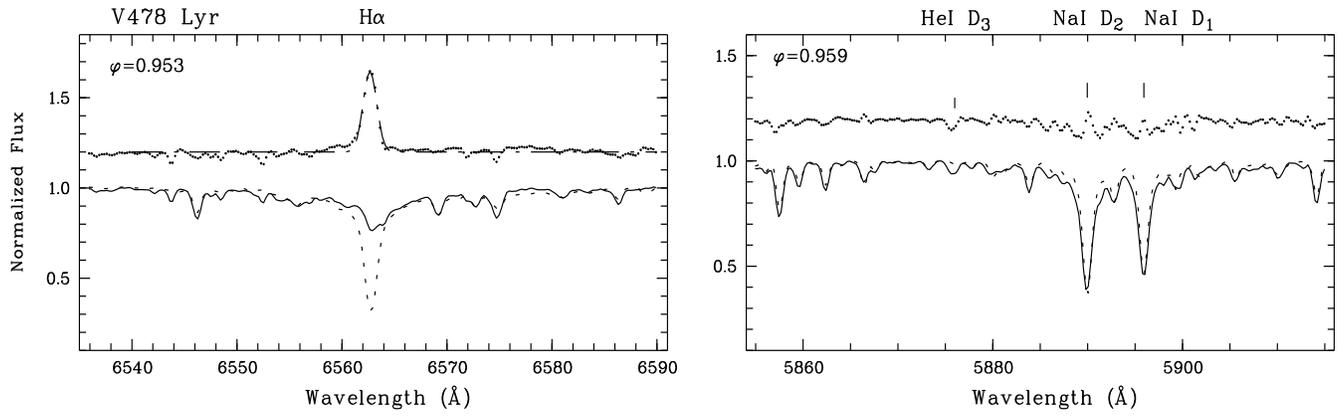,bbllx=288pt,bblly=73pt,bburx=450pt,bbury=627pt,height=5.6cm,width=18.0cm,angle=270,clip=}}
\caption[ ]{H$\alpha$, Na~{\sc i} D$_{1}$, D$_{2}$, and He~{\sc i} D$_{3}$ 
spectra of V478 Lyr
\label{fig:plot_v478lyr_hahe} }
\end{figure*}

\begin{figure*}
{\psscalefirst
\vspace{-0.5cm}
\hspace{-1.8cm}
{\psfig{figure=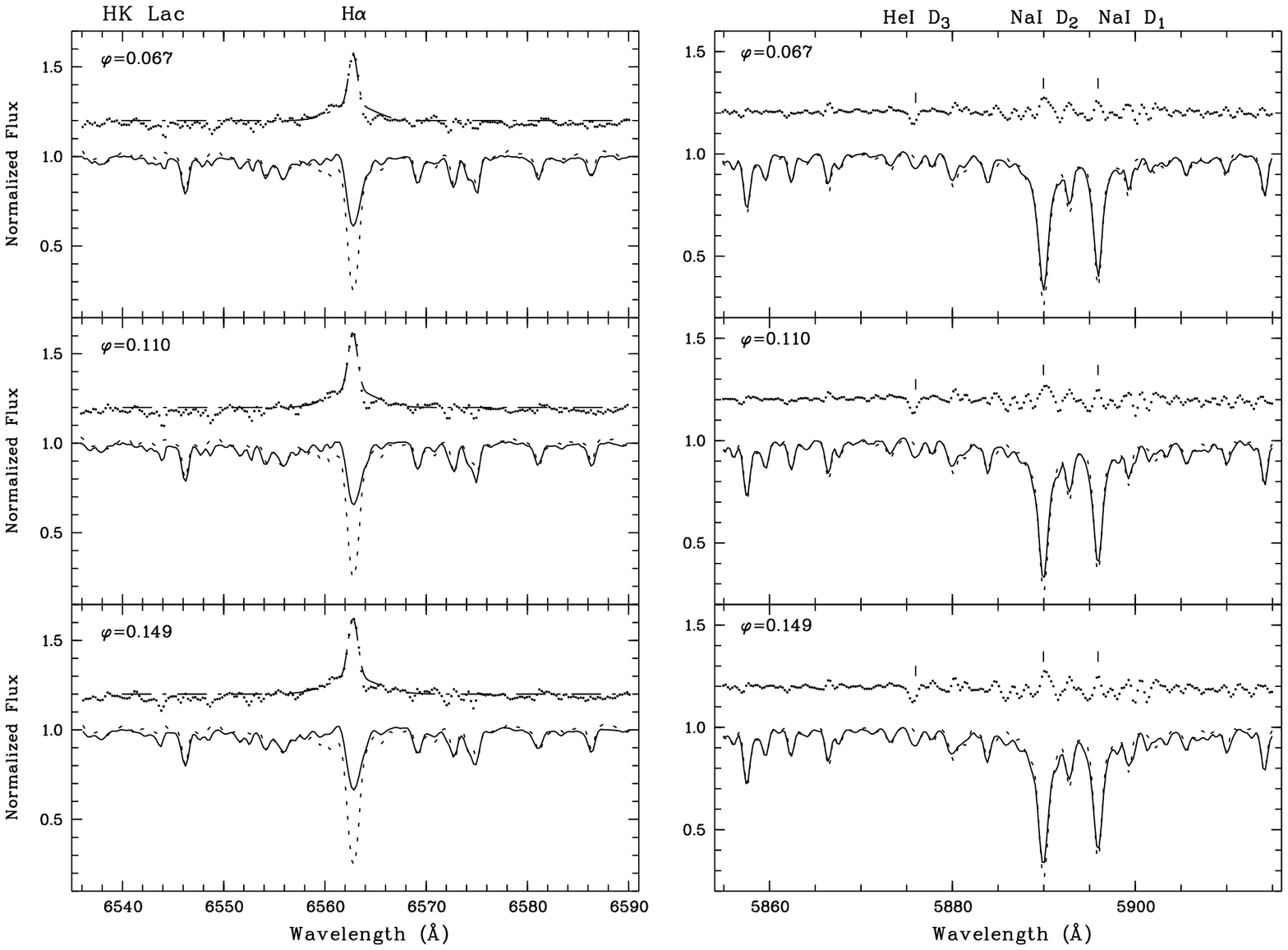,height=27.5cm,width=18.0cm,rheight=14.4cm,rwidth=6cm,angle=270}}
}
\caption[ ]{H$\alpha$, Na~{\sc i} D$_{1}$, D$_{2}$, and He~{\sc i} D$_{3}$ 
spectra of HK Lac
\label{fig:plot_hklac_hahe} }
\end{figure*}

\subsection{HD 12545 (XX Tri, BD +34 363) }

This extremely active RS CVn binary
is a single-lined spectroscopic binary of spectral type K0III.
It has very strong Ca~{\sc ii} H \& K and H$\epsilon$ emission lines 
and the H$\alpha$ line in emission above the continuum
(Strassmeier et al. 1990; Bopp et  al. 1993; Montes et al. 1995c).
This system shows the largest amplitude of light variation 
from spots yet recorded (Hampton et al. 1996).

Our observed H$\alpha$ spectrum at the orbital phase 0.401 
(Fig.~\ref{fig:plot_xxtri_hahe}) shows 
a strong and very broad H$\alpha$ emission above the continuum.
The synthesized spectrum has been constructed with a reference star of
spectral type K0III obtaining a satisfactory fit.
The subtracted spectrum exhibits an asymmetric profile
that is not well matched using a Gaussian fit, therefore
we fit the profile by means two Gaussian components.
In the Na~{\sc i} line region a clear excess emission in the 
D$_{1}$ and D$_{2}$ lines is present, in agreement with the strong
activity of this system.
The behaviour of the He~{\sc i} D$_{3}$ line is not clear in this spectrum,
but a weak absorption seems to appear.

\subsection{UX Ari (HD 21242)}

The description of the simultaneous H$\alpha$, Na~{\sc i} D$_{1}$, D$_{2}$,
and He~{\sc i} D$_{3}$  observations and the detection of a flare
on this system can be found in Montes et al. (1996b).

\subsection{V711 Tau (HR 1099, HD 22468)}

V711 Tau, one of the most active of the RS CVn binaries,
is a double-lined spectroscopic binary whose components have
spectral types G5IV and K1IV.
 Our Ca~{\sc ii} H \& K analysis of this system (FFMCC)
 showed that both components present
emissions. The cool component is the more active star
 in the system, and also presents the H$\epsilon$ line in emission.
This system shows the H$\alpha$ 
line in emission above the continuum and the spectral subtraction 
(Montes et al. 1995b) 
reveals that the K1 star is responsible for most of 
the excess H$\alpha$ emission.
Recently UV observations obtained with the HST's GHRS 
(Wood et al. 1996; Dempsey et al. 1996b, c) indicate that the transition region 
lines of V711 Tau are emitted almost entirely by the K1 star, 
and the G star contributes 14\% to the chromospheric Mg~{\sc ii} h \& k lines 
flux.

We report here five new observations of this systems in 
the H$\alpha$ line region at the orbital phases 0.92, 0.26, 0.28, 0.61 and 0.64
(1995 September 13-15) (see Fig.~\ref{fig:plot_v711tau_hahe})
that confirm the results obtained in our previous observations
taken in 1988, 1986 and 1992.
The H$\alpha$ subtracted profiles presents at all the epochs broad wings and
are not well matched using a single-Gaussian fit.
These profiles have therefore been fitted using two Gaussian components.
The parameters of the broad and narrow components used in the two-Gaussian fit
are given in table~\ref{tab:measures_nb} and the corresponding profiles are 
plotted in Fig.~\ref{fig:plot_v711tau_hahe_nb}.
In contrast to the behaviour found in other systems in which 
we have used a two-Gaussian fit, the broad component of V711 Tau dominates 
the line profile (the average contribution of the broad component
to the total H$\alpha$ EW is 77\%, whereas in other stars the contribution
ranges between 35\% and 63\%). 
Wood et al. (1996) and Dempsey et al. (1996c) 
also found this behaviour in the broad component of the 
chromospheric and transition region lines of V711 Tau in comparison with
the broad components found in the less active stars AU Mic and Capella.

In Fig.~\ref{fig:plot_v711tau_hahe_nb} we can also see that,
the contribution of the broad component is larger at orbital phases near 0.2.
(80-87\%) than at larger orbital phases (63-72\%)
and this behaviour remains at different epochs 
(see Fig.~\ref{fig:plot_v711tau_hahe_nb}).  
These changes indicate not only that the broad component is the result
of microflaring in the chromosphere, as in other stars, but that 
large and long-lived chromospheric flares,
that take place in this system (Dempsey et al. 1996c), could also produced
enhanced emission in the extended line wings.

Variable excess emission appears in the 
Na~{\sc i} D$_{1}$ and D$_{2}$ lines of the corresponding
subtracted spectra, however, no detectable absorption is observed
in the He~{\sc i} D$_{3}$ line.

\subsection{V833 Tau (HD 283750)}

V833 Tau is a BY Dra system (dK5e) and one of the hottest known
flare stars.
In our previous observations of this system we have found
a moderate H$\alpha$ emission above the continuum that
presents a variable little central self-reversal (Montes et al. 1995a, b) 
and a strong H$\beta$ excess emission (Montes et al. 1995d).

We analyse here three new observations of this systems in 
the H$\alpha$ line region at orbital phases 0.762, 0.319, and 0.851
(Fig.~\ref{fig:plot_v833tau_hahe})
that confirm the behaviour previously noted.
The self-reversal feature changes its wavelength-position
across the emission H$\alpha$ line profile, however, it appears always centered
with the corresponding H$\alpha$ absorption line of the synthesized spectrum.

The Na~{\sc i} D$_{1}$ and D$_{2}$ lines exhibit
very broad wings in the observed spectra, 
as correspond to a star of spectral type as later as K5.
Although the observed and synthesized spectra
are not well matches, due to problems in the normalization,
a clear excess emission in the D$_{1}$ and D$_{2}$ lines can be 
seen in the subtracted spectra.
The He~{\sc i} D$_{3}$ line is not detected in this star.

\subsection{V1149 Ori (HD 37824)}

A single-lined spectroscopic binary classified as K1III.
Our previous observations 
reveal a clear excess H$\alpha$ emission (Montes et al. 1995a, b), 
strong Ca~{\sc ii} H \& K and H$\epsilon$ emission lines 
(Montes et al. 1995c). 

We present here one spectrum at the orbital phase 0.439
(see Fig.~\ref{fig:plot_v1149ori_hahe}).
The difference with respect to a K0III reference star reveals a 
lower excess H$\alpha$ emission than the obtained in our previous
observations in 1992 (Montes et al. 1995a).
The narrow absorption that appears in the red wing of the excess H$\alpha$
emission in the subtracted spectra could be attributed to a telluric line.

The spectral subtraction reveals that this star has no
measurable filling-in of the Na~{\sc i} D$_{1}$ and D$_{2}$ lines.
However, a clear absorption in the He~{\sc i} D$_{3}$ line 
appears in the subtracted spectrum.

\subsection{MM Her (HD 341475)}

Double-lined spectroscopic binary (G2/K0IV) with partial eclipses.
This system has
Ca~{\sc ii} H \& K emission lines from both components and 
excess H$\alpha$ emission from the cool component 
(FFMCC, Montes et al. 1995a).

We have obtained three new spectra (Fig.~\ref{fig:plot_mmher_hahe}) 
of this system
in the H$\alpha$ and Na~{\sc i} D$_{1}$ and D$_{2}$ line regions
at orbital phases 0.498, 0.630, and 0.745.
At phase 0.498 the hot component hides a 0.30 fraction of the cool one,
(see the diagram of Fig.~\ref{fig:plot_mmher_hahe}) 
and the absorption lines from both
components appear overlapped in the observed spectrum.
However, at phases 0.630 and 0.745 there is not eclipse and
the lines are clearly wavelength-shifted.
The more intense H$\alpha$ absorption observed in these spectra
corresponds to the hot component and the less intense and blue-shifted 
absorption corresponds to the cool one.
The spectral subtraction,
using  G2IV and K0IV as reference stars and a relative contribution 
of 0.5/0.5, indicates that the excess H$\alpha$ 
emission arises only from the cool component.
The excess H$\alpha$ emission line profiles obtained are 
well matched using a two-component Gaussian fit.

In the subtracted spectra of
the Na~{\sc i} D$_{1}$ and D$_{2}$ lines region we can also see 
an excess emission in these lines from
the cool component.
The narrow absorption that appears near the position of the hot component
could be due to deficient sky correction in these spectra.
The He~{\sc i} D$_{3}$ line is not present.

\subsection{V815 Her (HD 166181)}

Single-lined spectroscopic binary.
Our previous Ca~{\sc ii} H \& K and H$\alpha$ observations 
(FFMCC, Montes et al. 1995a) indicate that 
the hot star is the active component.
Multiwavelength observations of this system have been
recently reported by Dempsey et al. (1996a).

We have taken three spectra of this system at orbital phases 
0.978, 0.520, 0.099 (see Fig.~\ref{fig:plot_v815her_hahe}).
In the H$\alpha$ line region the spectra show a filling-in absorption line
with noticeable night to night changes.
The excess H$\alpha$ emission obtained with the spectral subtraction of a G5~V 
reference star is larger at the orbital phases near to 0.0.
The subtracted H$\alpha$ profile presents broad wings and is
well matched using a fit with two Gaussian components (narrow and broad).
The narrow component is more important when the excess H$\alpha$ emission is
larger (see table~\ref{tab:measures_nb}).

A small filling-in is observed in the Na~{\sc i} D$_{1}$ and D$_{2}$ lines
at the three orbital phases, and only at phase 0.523 absorption in detected
in the He~{\sc i} D$_{3}$ line.

\subsection{BY Dra (HD 234677)}

The prototype of the BY$~$Dra stars.
Our previous Ca~{\sc ii} H \& K observations of this system (FFMCC)
clearly show that both components are
active with the hot component having the stronger Ca~{\sc ii} emission.
The two components also show  H$\epsilon$  in emission.

We present in this paper simultaneous 
H$\alpha$ and Na~{\sc i} D$_{1}$ and D$_{2}$ 
observations of this system at orbital phases 0.684 and 0.839
(see Fig.~\ref{fig:plot_bydra_hahe}). 
These spectra show strong H$\alpha$ emission above the continuum
and the Na~{\sc i} D$_{1}$ and D$_{2}$ lines with very broad
wings corresponding to the later spectral type of the components
of this system (K4V/K7.5V).

By applying the spectral subtraction technique 
we have obtained an asymmetric
excess H$\alpha$ emission line profile which has contributions from
both components. 
A two-Gaussian fit has been used 
to deblend the contribution of the hot and cool components 
to the line profile.
This fit reveals that the hot component have the stronger excess 
H$\alpha$ emission EW in agreement with the behaviour observed by
us in the Ca~{\sc ii} H \& K lines.

The subtracted spectra in the 
Na~{\sc i} D$_{1}$ and D$_{2}$ line region reveal
that the excess emission in these lines also arises from both components.
The He~{\sc i} D$_{3}$ line is not detected in these spectra.

\subsection{V775 Her (HD 175742) }

Single-lined spectroscopic binary (K0V/[K5-M2V]) 
with strong Ca~{\sc ii} H \& K emission lines 
from the hot component (FFMCC).
The H$\alpha$ feature may change from a weak absorption feature to emission
above the continuum on times scales of hours (Xuefu and Huisong 1984).

In our H$\alpha$ spectrum at the orbital phase 0.394 
(Fig.~\ref{fig:plot_v775her_hahe}) we can see the H$\alpha$
line of the hot component totally filled-in by emission 
and a small emission bump red-shifted
in relation to the absorption lines.
By subtracting the synthesized spectrum,
constructed with a K0IV reference star, we have obtained
a strong excess H$\alpha$ emission coming from the hot component
and a small excess emission, red-shifted 1.9 \AA$\ $ with respect to the 
emission of the hot component.
 This small excess  perhaps could be attributed to
the cooler star of the system, whose assumed spectral type is
K5-M2V and whose contribution to the observed spectra is negligible.

In the Na~{\sc i} lines region the spectral subtraction points out 
a filling-in of the D$_{1}$ and D$_{2}$ lines and not 
detectable absorption in the He~{\sc i} D$_{3}$ line.

\subsection{V478 Lyr (HD 178450) }

This BY$~$Dra system is a
single-lined spectroscopic binary with
strong Ca~{\sc ii} H \& K emissions from the hot component (FFMCC)
and a filled-in H$\alpha$ absorption line (Fekel 1988).

The H$\alpha$ spectrum of this system exhibits 
a strong filling-in absorption line
(see Fig.~\ref{fig:plot_v478lyr_hahe}). 
By subtracting the synthesized spectrum constructed with a
G8V star we have obtained strong excess H$\alpha$ emission,
a small excess emission in the Na~{\sc i} D$_{1}$ and D$_{2}$ lines
and a clear absorption in the He~{\sc i} D$_{3}$ line.

\subsection{HK Lac (HD 209813)}

HK~Lac is a single-lined spectroscopic binary (F1V/K0III) with very
strong  Ca~{\sc ii} H \& K  emissions, the H$\epsilon$ line in emission
and an important excess H$\alpha$ emission (FFMCC; Montes el at. 1995a).

This system shows a very variable H$\alpha$ profile (from filled-in absorption 
to moderate emission) and flares (see Catalano \& Frasca 1994). 
However, in our six H$\alpha$ spectra taken in three consecutive nights, 
with orbital phases from 0.067 to 0.153, we always observe
filled-in absorption line with small night to night variations in the
excess H$\alpha$ emission from the cool component.
In Fig.~\ref{fig:plot_hklac_hahe} a spectrum of each night is showed.

The observed spectra are well matched using a K0III as reference star.
The subtracted spectra show
an important excess emission in the Na~{\sc i} D$_{1}$ and D$_{2}$ lines
and a clear absorption in the He~{\sc i} D$_{3}$ line.

\begin{figure*}
\hbox{
{\psfig{figure=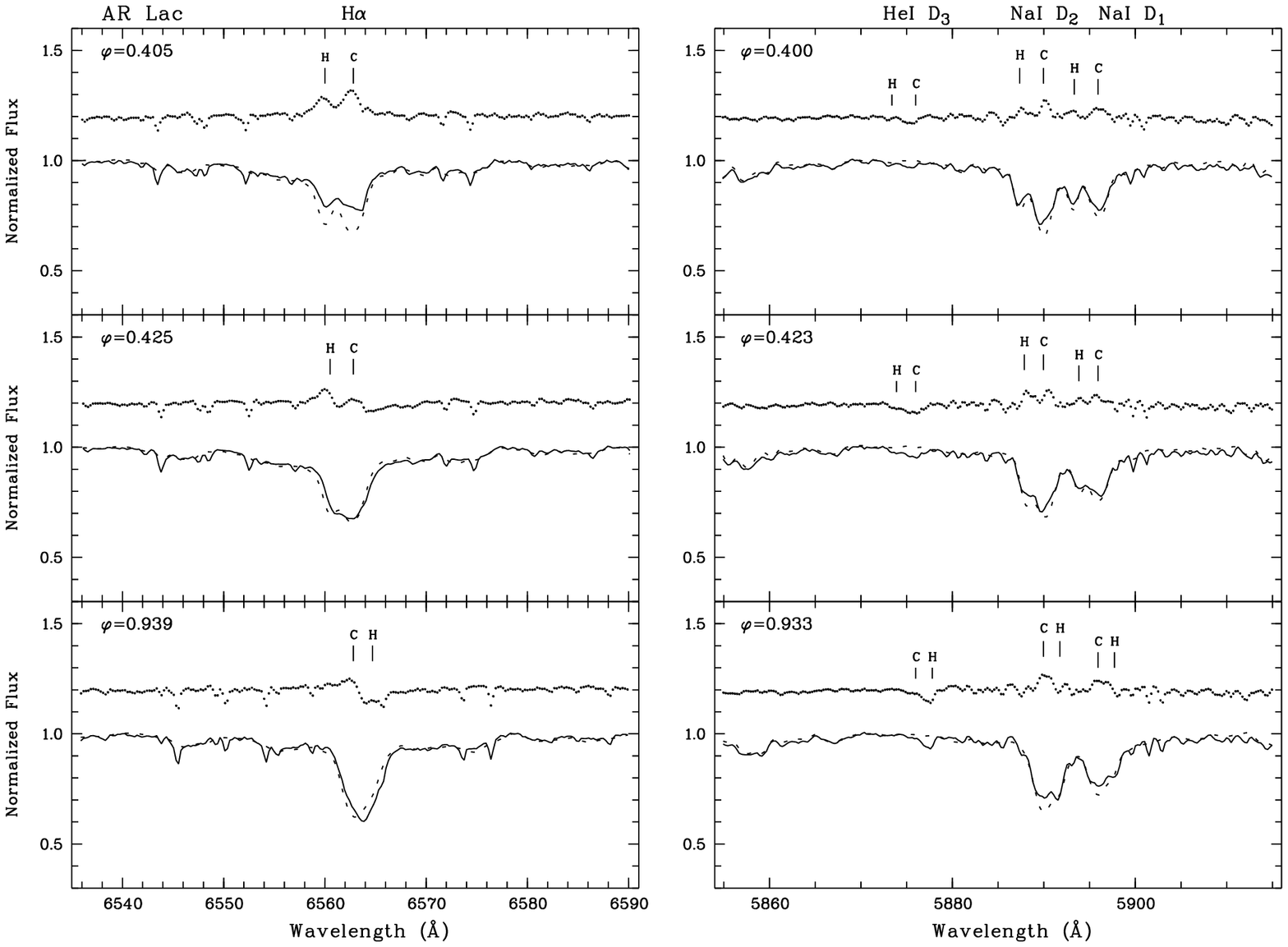,bbllx=44pt,bblly=73pt,bburx=450pt,bbury=627pt,height=20.5cm,width=18.0cm,rheight=20.5cm,angle=270,clip=}}
\hspace{-16.9cm}
\vbox{
{\psfig{figure=ds5809f15_a.ps,bbllx=85pt,bblly=85pt,bburx=344pt,bbury=202pt,height=1.8cm,width=4.0cm,clip=}}
\vspace{4.2cm}
{\psfig{figure=ds5809f15_b.ps,bbllx=85pt,bblly=85pt,bburx=344pt,bbury=202pt,height=1.8cm,width=4.0cm,clip=}}
\vspace{4.2cm}
{\psfig{figure=ds5809f15_c.ps,bbllx=85pt,bblly=85pt,bburx=344pt,bbury=202pt,height=1.8cm,width=4.0cm,clip=}}
\vspace{1.3cm}
}
}
\caption[ ]{H$\alpha$, Na~{\sc i} D$_{1}$, D$_{2}$, and He~{\sc i} D$_{3}$ 
spectra of AR Lac
\label{fig:plot_arlac_hahe} }
\end{figure*}

\begin{figure*}
{\psfig{figure=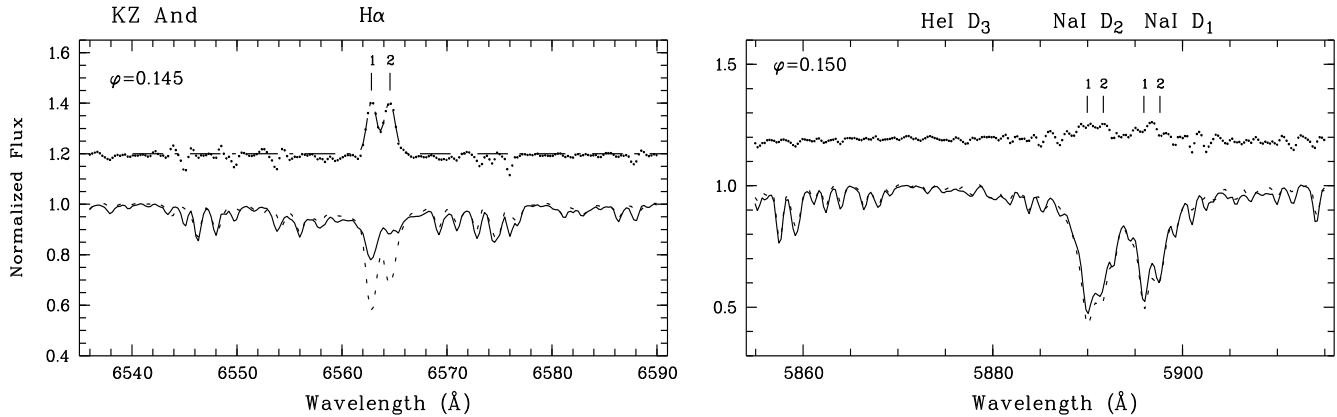,bbllx=288pt,bblly=73pt,bburx=450pt,bbury=627pt,height=5.6cm,width=18.0cm,angle=270,clip=}}
\caption[ ]{H$\alpha$, Na~{\sc i} D$_{1}$, D$_{2}$, and He~{\sc i} D$_{3}$ 
spectra of KZ And 
\label{fig:plot_kzand_hahe} }
\end{figure*}

\begin{figure*}
{\psfig{figure=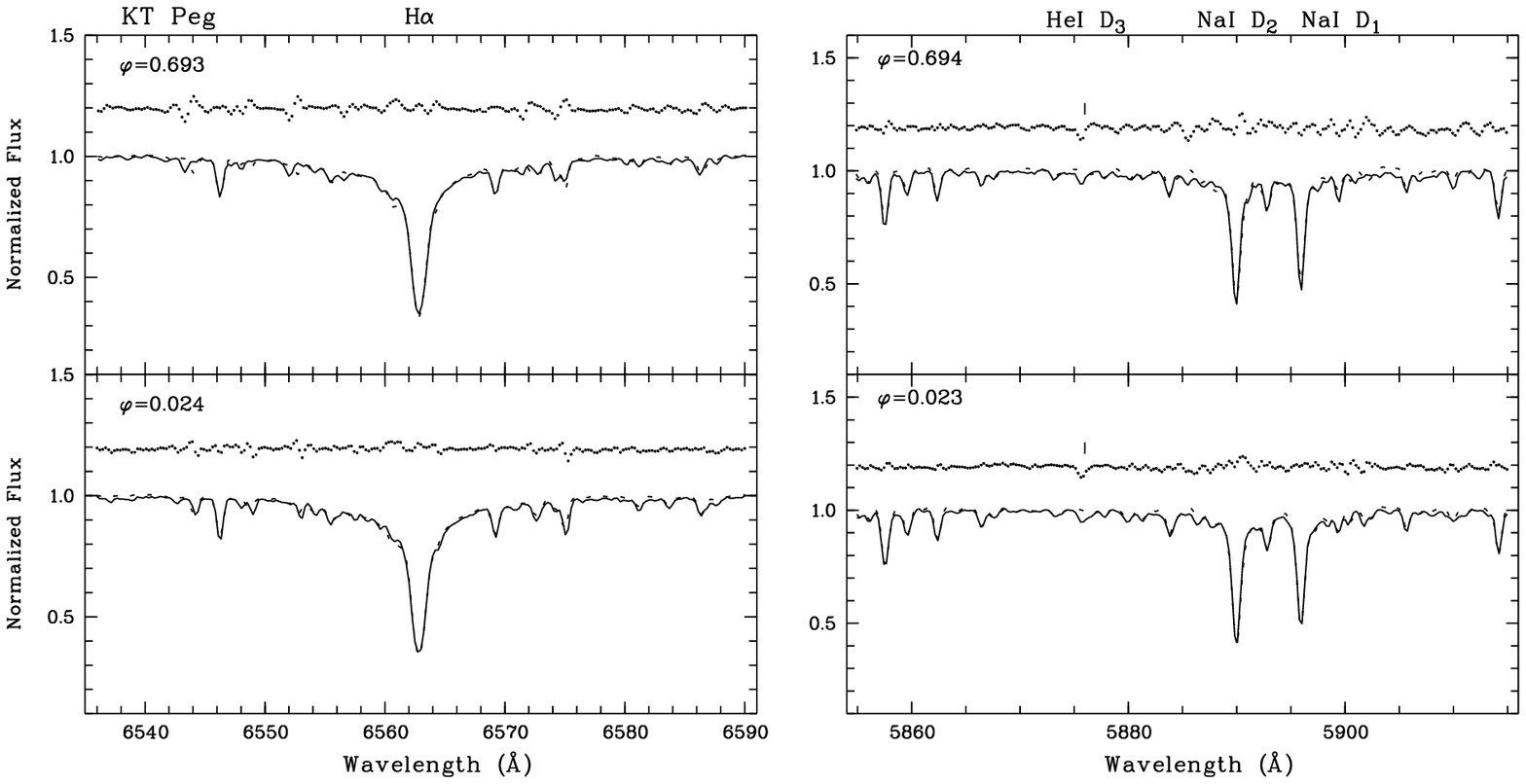,bbllx=44pt,bblly=73pt,bburx=329pt,bbury=627pt,height=10.0cm,width=18.0cm,angle=270,clip=}}
\caption[ ]{H$\alpha$, Na~{\sc i} D$_{1}$, D$_{2}$, and He~{\sc i} D$_{3}$ 
spectra of KT Peg
\label{fig:plot_ktpeg_hahe} }
\end{figure*}

\begin{figure*}
{\psfig{figure=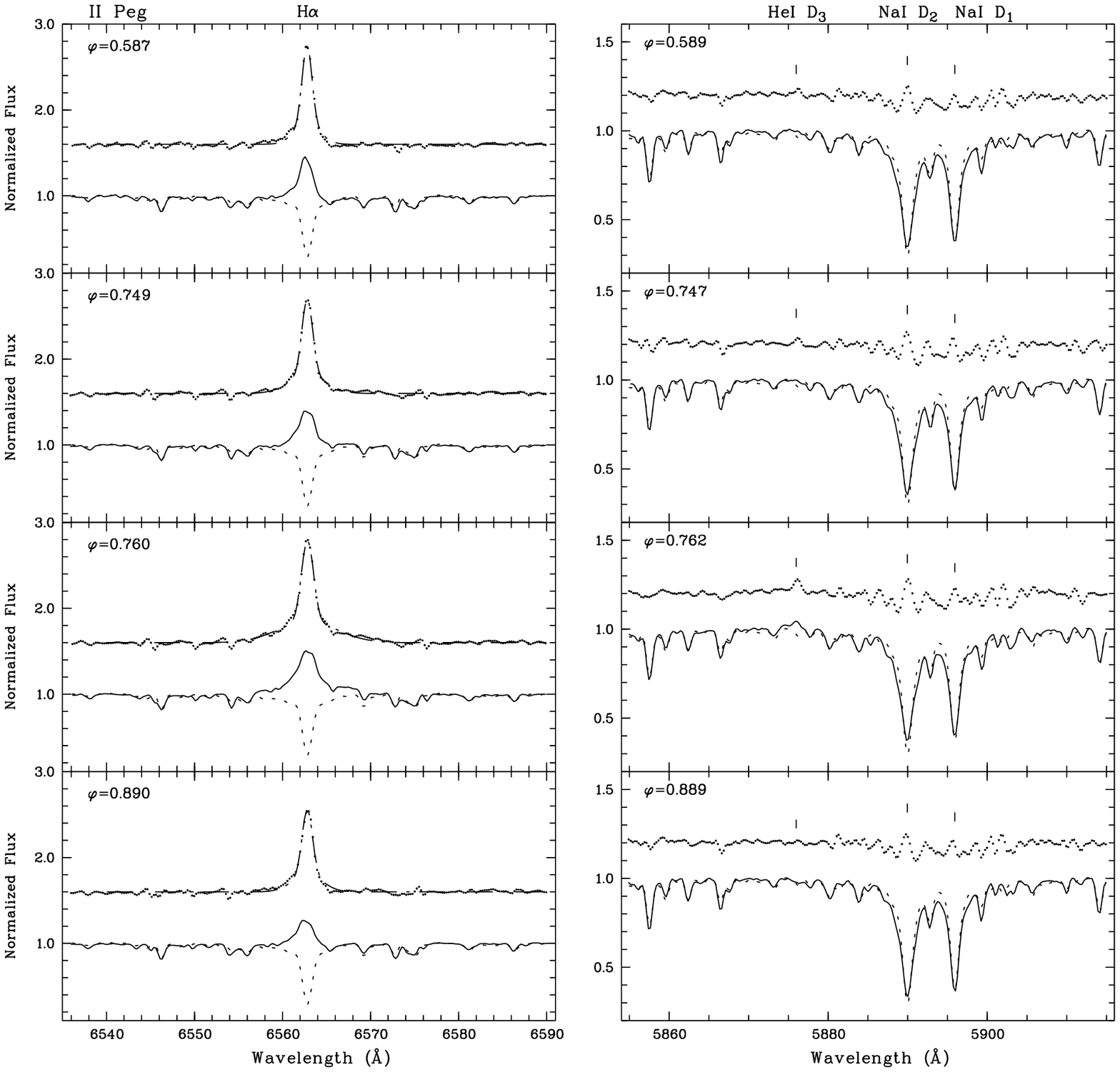,bbllx=44pt,bblly=73pt,bburx=570pt,bbury=627pt,height=19.0cm,width=18.0cm,angle=270,clip=}}
\caption[ ]{H$\alpha$, Na~{\sc i} D$_{1}$, D$_{2}$, and He~{\sc i} D$_{3}$ 
spectra of II Peg
\label{fig:plot_iipeg_hahe} }
\end{figure*}

\subsection{AR Lac (HD 210334)}

The total eclipsing binary AR Lac is one of the
best studied RS$~$CVn systems.
Both components of the system (G2IV/K0IV) are active (CABS), 
however, due to the orbital phase, in our previous 
Ca~{\sc ii} H \& K and H$\alpha$ observations (FFMCC)
it was not possible to separate the contribution from each component.

We analyse here H$\alpha$ observations of this system at three orbital phases.
In Fig.~\ref{fig:plot_arlac_hahe} we have superimposed to each spectrum
the corresponding geometrical position of the cool and hot components.
At orbital phase 0.405 there is not eclipse and the subtracted spectrum
exhibits excess H$\alpha$ emission from both components, being a slightly 
larger that corresponding to the cool one.
At phase 0.425 the hot component hides a 4~$\%$ of the cool one and
the excess emission obtained is now slightly larger in the hot component.
This change in the excess emission observed 
in both components could be attributed 
to the hot component hiding of prominence-like material or other active 
regions responsible for the H$\alpha$ emission of the cool component.

In the observation at the orbital phase 0.939, when a 33~$\%$  
of the hot component is hidden by the cool component, 
we detect an excess H$\alpha$ absorption
located at the wavelength position corresponding to the hot component.
This excess absorption indicates the presence of prominence-like
extended material seen off the limb of the cool component that absorbs the 
photospheric continuum radiation of the star behind.
Similar prominence-like structures have been found in other eclipsing RS CVn 
systems (see Hall \& Ramsey 1992) and recently, Siarkowski et al. (1996)
have found geometrically extended structures in the corona of AR Lac.

At these three orbital phases the Na~{\sc i} D$_{1}$ and D$_{2}$
lines also present excess emissions from both components and the
He~{\sc i} D$_{3}$ shows absorption features in the subtracted spectra.
At orbital phase 0.933 the He~{\sc i} D$_{3}$ line presents
a large excess absorption at the wavelength position corresponding to the hot 
component, which could also be attributed to the prominence-like material.

\begin{figure}
{\psfig{figure=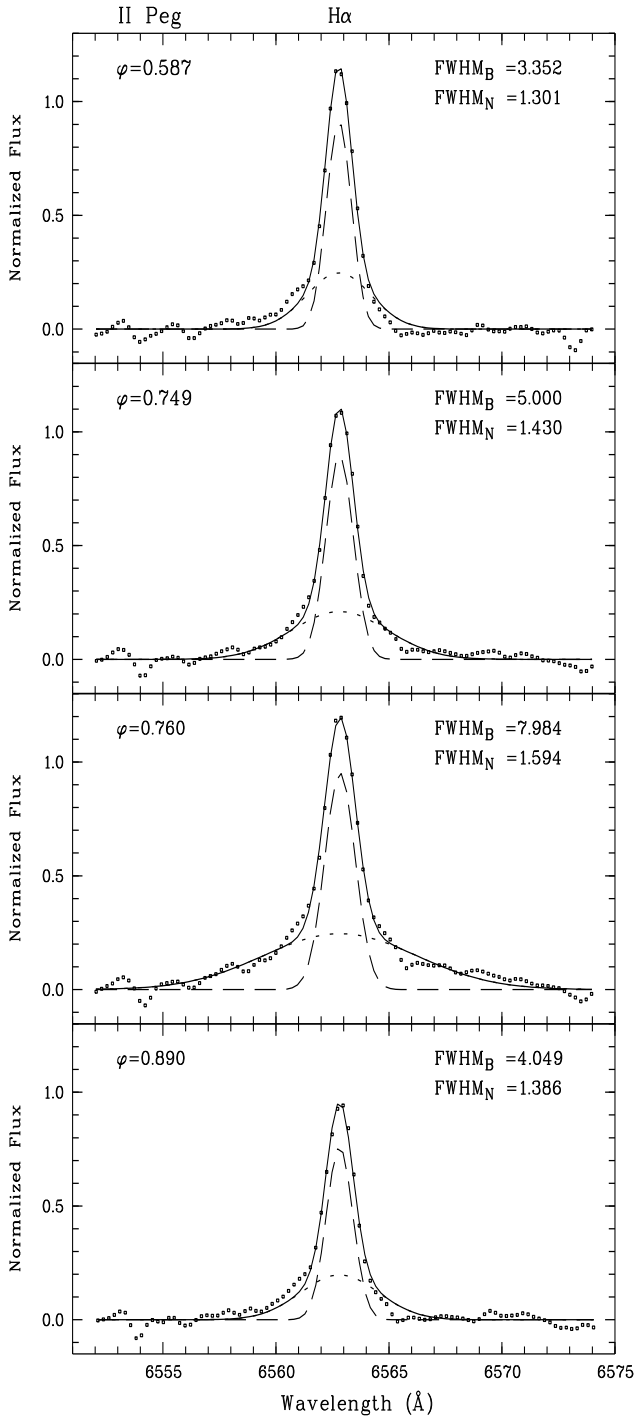,bbllx=44pt,bblly=73pt,bburx=570pt,bbury=360pt,height=19.0cm,width=8.6cm,angle=270,clip=}}
\caption[ ]{Subtracted H$\alpha$ profiles of II Peg  
at different orbital phases (dotted line).
We have superposed the two Gaussian components fit (solid-line).
The sort-dashed-line represents the broad component 
and the large-dashed-line the narrow one
\label{fig:plot_iipeg_hahe_nb} }
\end{figure}

\subsection{KZ And (HD 218738)}

KZ And is the component B of the visual pair ADS 16557, and is a double-lined
spectroscopic binary with spectral types dK2/dK2.
This system presents Ca~{\sc ii} H \& K and H$\epsilon$ emissions
and excess H$\beta$ emission from both components 
(FFMCC, Montes et al. 1995c, d).

By applying the spectral subtraction technique 
we have found that also both components present
excess emission in the H$\alpha$ line and in the
in the Na~{\sc i} D$_{1}$ and D$_{2}$ lines, with very similar intensities
(see Fig.~\ref{fig:plot_kzand_hahe}).
The He~{\sc i} D$_{3}$ line is not present in the subtracted spectrum.
The synthesized spectrum has been constructed with two K2V stars
and with a contribution of each
component to the total continuum of 0.58/0.42.

%

\subsection{KT Peg (HD 222317) }

This system is a double-lined spectroscopic binary (G5V/K6V) with
small Ca~{\sc ii} H \& K emissions from both components (Montes et al. 1995c).

We analyse here two spectra at the orbital phases 0.693 and 0.024
(see Fig.~\ref{fig:plot_ktpeg_hahe}).
We have constructed the synthesized spectrum with two reference stars of
spectral types G2IV and K3V taking into account 
that the hot component has the larger contribution to the
continuum. 
In both subtracted spectra no detectable excess emission is observed in the
H$\alpha$ line nor in the Na~{\sc i} D$_{1}$ and D$_{2}$ lines.
The He~{\sc i} D$_{3}$ line appears in absorption.

\begin{figure*}
{\psfig{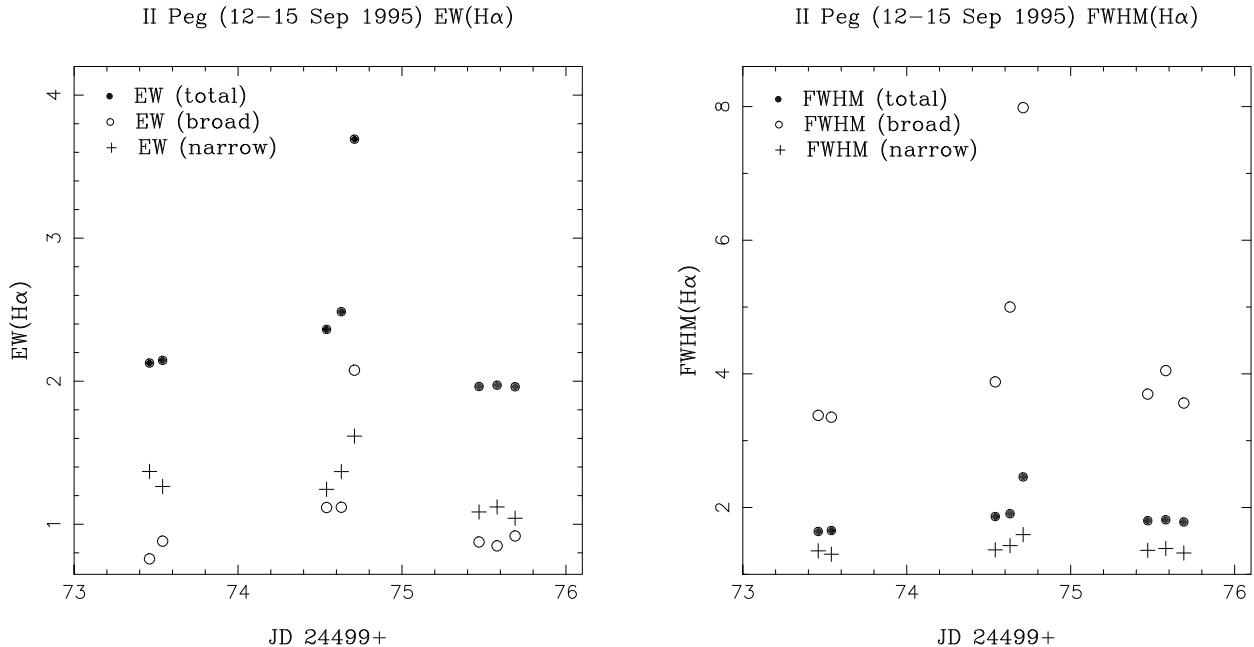}}
\caption[ ]{Changes in the EW(H$\alpha$) and FWHM(H$\alpha$)
of II Peg from 12-15 September 1995. Note the different behaviour of the
broad and narrow components
\label{fig:iipeg_ew_fwhm} }
\end{figure*}

\subsection{II Peg (HD 224085) }

This is a single-lined spectroscopic binary (K2-3V-IV) with strong Ca~{\sc ii} H \& K
emission lines and variable H$\alpha$ emission above the continuum.
We have reported a strong excess emission in the H$\beta$ line
(Montes et al. 1995d).

We present here eight spectra of this system taken in three consecutive nights 
and with orbital phases ranging from 0.58 to 0.91.
The strong H$\alpha$ emission present in the subtracted spectra 
in all the phases shows remarkable night to night changes 
(see Fig.~\ref{fig:plot_iipeg_hahe}). In the second
night the spectrum at orbital phase 0.760 
 shows a remarkable excess H$\alpha$ emission
enhancement with regard to the other phases,  
which is much more noticeable in comparison with the spectra
of the first and third nights.
The excess H$\alpha$ emission EW increases in a factor of 1.9
in an interval of 1 day.
This fact suggests that we have detected a flare since enhancements
of this amount during flares are typical of chromospheric emission lines
(Simon et al. 1980, Catalano \& Frasca 1994) and has also been observed
by us in the flare detected in UX Ari (Montes et al. 1996b).

The He~{\sc i} D$_{3}$ line appears in emission in the observed spectrum
only at orbital phase 0.760 (see Fig.~\ref{fig:plot_iipeg_hahe})
corresponding to the increase of the emission observed in the H$\alpha$ line.
This fact supports the detection of a flare-like event, since in the Sun
the He~{\sc i} D$_{3}$ line appears as absorption in plage and weak flares
and as emission in strong flares (Zirin 1988).
At the other orbital phases of the first and second night an small emission
in He~{\sc i} D$_{3}$ is also present in the subtracted spectrum. 
However, in the third night where the H$\alpha$ EW present the lower values
the He~{\sc i} D$_{3}$ emission is not observed.

The subtracted spectra show that
the H$\alpha$ profile presents broad wings, which are
much more remarkable in the flare spectrum.
In Fig.~ \ref{fig:plot_iipeg_hahe_nb} we have represented
the subtracted spectrum at phases 0.587, 0.749, 0.760 and 0.890 
and the corresponding narrow and broad components used to perform 
the two-component Gaussian fit.

The changes in the EW(H$\alpha$) and FWHM(H$\alpha$) of the total subtracted 
spectra and of the corresponding narrow and broad components
during the three nights can be seen in 
Fig.~\ref{fig:iipeg_ew_fwhm}. Note the strong change in the FWHM of the broad
component during the flare.

\begin{figure*}
{\psfig{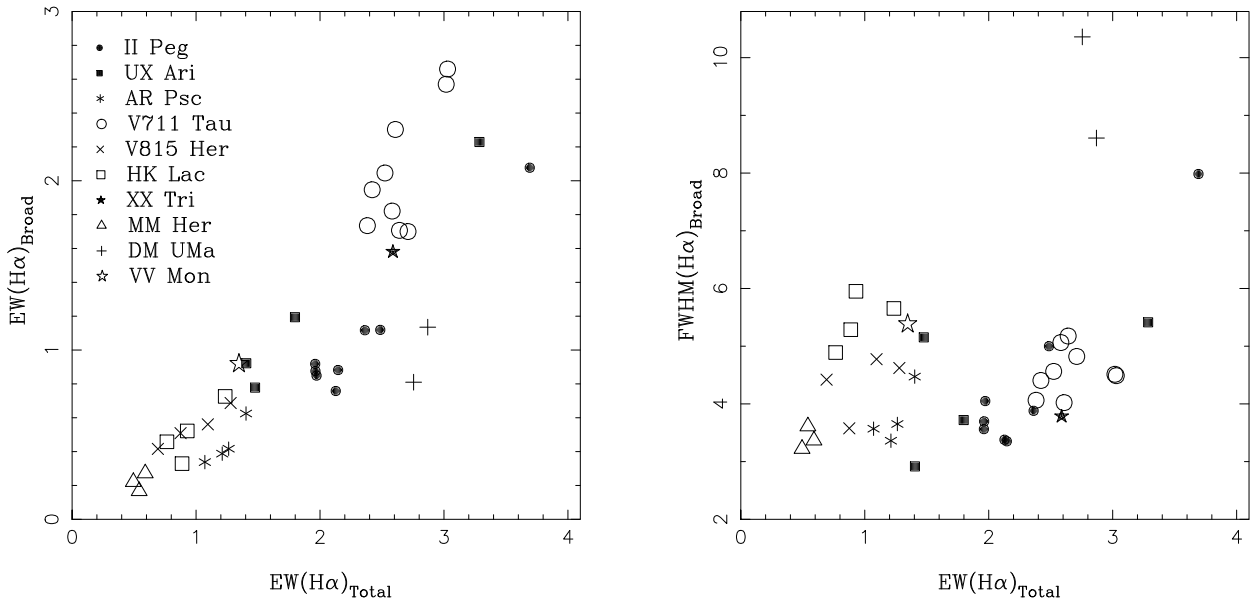}}
\caption[ ]{EW(H$\alpha$) (left-panel) and  FWHM(H$\alpha$) (right-panel)
of the broad component versus the total EW(H$\alpha$).
We have used different symbols for each star
\label{fig:fwhm_b_ew_t} }
\end{figure*}

\section{Discussion}

\subsection{The excess H$\alpha$ emission}


The inspection of the subtracted H$\alpha$ spectra shows that 
in some stars the emission line profile has very broad wings,
and is not well matched using a single-Gaussian fit.
These profiles have therefore been fitted using 
two Gaussian components: a narrow component having a FWHM of
45-90 km~s$^-1$ and a broad component with a FWHM ranging from
133 to 470 km~s$^-1$.
In Table~\ref{tab:measures_nb} we list the parameters (I, FWHM, EW)
of the broad and narrow components.
As can be seen in this table the average contribution of the broad component
to the total EW of the line ranges from 35\% in AR Psc to 77\% in  V711 Tau.
We have observed this behaviour in the chromospheric H$\alpha$ line 
only in the most active systems of the sample, the stars with 
intense H$\alpha$ emission above the continuum 
(AR Psc, XX Tri, UX Ari, V711 Tau (Fig.~\ref{fig:plot_v711tau_hahe_nb}), 
II Peg (Fig.~\ref{fig:plot_iipeg_hahe_nb})) and also 
in systems with important excess H$\alpha$ emission
(MM Her, V815 Her, HK Lac).
Furthermore, a revision of the H$\alpha$ spectra of a large sample 
of chromospherically active binaries previously analysed by us 
(Montes et al. 1995a, b) indicates that 
the very active systems DM UMa and VV Mon also have broad components and 
that some systems also studied in this paper 
(V711 Tau, V815 Her, HK Lac) exhibit this behaviour at different epochs.

This parameterization of the subtracted H$\alpha$ profile 
using a narrow and a broad component have been only reported 
until now for the RS CVn system DM UMa by Hatzes (1995) who suggests that  
the broad component could result from large-scale motions or winds in the
chromosphere.
Similar broad components have also been found
in several transition region lines of
the dM0e star AU Mic and the RS CVn systems
Capella and V711 Tau using
high-resolution UV observations obtained with the HST's GHRS
(Linsky \& Wood 1994; Linsky et al. 1995;
Wood et al. 1996; Dempsey et al. 1996b, c; Robinson et al. 1996).
The broad components in the transition region lines are interpreted
by Linsky \& Wood (1994) as arising from microflaring, because
these broad profiles are reminiscent of the broad profiles observed
in solar transition region explosive events, which are thought to be
associated with emerging magnetic flux regions where field
reconnection occurs. 

The microflares are frequent, short-duration, energetically
weak disturbances, i.e. they are the low-energy extension of flares,
and therefore have large-scale motions associated that could
explain the broad wings observed in these lines.
The microflaring activity could occur not only in the transition region
but also in the chromosphere of very active stars as indicates 
the detection of broad components
in the chromospheric Mg~{\sc ii} h \& k lines of V711 Tau (Wood et al. 1996)
but not in the chromospheric lines of the less active star Capella 
(Linsky et al. 1995).

Our detection of broad wings in the chromospheric H$\alpha$ line of the
most active systems of our sample allows us to 
conclude that microflaring occurs also in
the chromosphere and that it is much more important in extremely active stars.
Furthermore, within the group of stars that present this phenomenon
a correlation between the contribution of the broad components 
and the degree of stellar activity seems to be 
present, as can be seen in Fig.~\ref{fig:fwhm_b_ew_t} left-panel where 
we have plotted for each star the EW of the broad component
versus the total excess H$\alpha$ EW.
This correlation was also noted by
Wood et al. (1996) when compared the dominant broad component 
of V711 Tau with the smaller broad component of the less active 
stars AU Mic and Capella.

On the other hand, when we plot the FWHM of the broad component
 versus the total H$\alpha$ EW
(Fig.~\ref{fig:fwhm_b_ew_t} right-panel)
 a general trend is not observed.
However, in this case the relation appears for individual star,
i.e. there is an increase in the FWHM when the activity
level increases, which
 is consistent with the hypothesis
that microflaring is responsible for the broad component emission.


We have also found that the larger changes in the excess 
H$\alpha$ emission in the stars analyzed appear to occur 
predominantly in the broad component, 
as have been already noted in the case of DM UMa by 
Hatzes (1995).
An extreme case of this behaviour is the strong change
in the broad component that occurs
during the flares detected in UX Ari and II Peg
(see Fig.~\ref{fig:iipeg_ew_fwhm} and \ref{fig:fwhm_b_ew_t}).
Since large scale mass motions do occur in solar flares, 
a large flare in
these two systems may explain the increase of the H$\alpha$ emission 
and the very broad wings observed in these spectra.

\begin{table*}
\caption[]{Parameters of the broad and narrow Gaussian components
used in the fit of the H$\alpha$ subtracted spectra of the stars 
analysed in this paper and in V711 Tau, V815 Her, HK Lac, DM UMa, and VV Mon
from our previous observations (Montes et al. 1995a, b)
\label{tab:measures_nb}}
\begin{flushleft}
\scriptsize
\begin{tabular}{lcccccccccc}
\noalign{\smallskip}
\hline
\noalign{\smallskip}
\noalign{\smallskip}
        &    &  
\multicolumn{4}{c}{H$\alpha$ broad component} &\ &
\multicolumn{4}{c}{H$\alpha$ narrow component} \\
\cline{3-6}\cline{8-11}
\noalign{\smallskip}
 {Name} & {$\varphi$} & 
{I} & FWHM & EW$_{\rm B}$ & EW$_{\rm B}$/EW$_{\rm T}$ & &  
{I} & FWHM & EW$_{\rm N}$ & EW$_{\rm N}$/EW$_{\rm T}$ \\
        &           & 
    & {\scriptsize (\AA)} & {\scriptsize (\AA)} & (\%) & & 
    & {\scriptsize (\AA)} & {\scriptsize (\AA)} & (\%) \\
\noalign{\smallskip}
\hline
\noalign{\smallskip}
AR Psc & 0.373 & 0.089 & 3.570 & 0.338 & 31.6 & & 0.541 & 1.272 & 0.732 & 68.3 \\
       & 0.443 & 0.132 & 4.473 & 0.625 & 44.5 & & 0.534 & 1.358 & 0.772 & 55.0 \\
       & 0.519 & 0.108 & 3.362 & 0.388 & 32.0 & & 0.587 & 1.319 & 0.823 & 68.0 \\
       & 0.524 & 0.107 & 3.652 & 0.417 & 33.0 & & 0.603 & 1.316 & 0.845 & 67.0 \\
\noalign{\smallskip}
XX Tri & 0.401 & 0.392 & 3.785 & 1.581 & 60.8 & & 0.712 & 1.132 & 1.007 & 38.7 \\
\noalign{\smallskip}
UX Ari & 0.419 & 0.302 & 3.720 & 1.194 & 66.4 & & 0.431 & 1.317 & 0.605 & 33.6 \\
       & 0.438 & 0.297 & 2.917 & 0.922 & 65.6 & & 0.344 & 1.316 & 0.483 & 34.4 \\
       & 0.576 & 0.145 & 5.153 & 0.778 & 50.8 & & 0.468 & 1.512 & 0.753 & 49.2 \\
       & 0.736 & 0.405 &{\bf 5.414}& 2.229 & 67.8 & & 0.580 & 1.723 & 1.069 & 32.4 \\
\noalign{\smallskip}
V711 Tau (1995) & 0.922 & 0.335 & 4.822 & 1.700 & 62.7 & & 0.502 & 1.852 & 0.990 & 36.5 \\
                & 0.261 & 0.425 & 4.563 & 2.046 & 81.1 & & 0.344 & 1.254 & 0.459 & 18.2 \\
                & 0.280 & 0.417 & 4.406 & 1.947 & 80.4 & & 0.344 & 1.265 & 0.463 & 19.1 \\
                & 0.606 & 0.318 & 5.176 & 1.707 & 64.6 & & 0.509 & 1.635 & 0.887 & 33.6 \\
                & 0.641 & 0.402 & 4.063 & 1.735 & 72.8 & & 0.406 & 1.486 & 0.642 & 27.0 \\
V711 Tau (1992) & 0.130 & 0.546 & 4.022 & 2.303 & 88.3 & & 0.234 & 1.226 & 0.305 & 11.7 \\
V711 Tau (1986) & 0.200 & 0.570 & 4.489 & 2.660 & 87.8 & & 0.251 & 1.381 & 0.369 & 12.2 \\
                & 0.260 & 0.576 & 4.513 & 2.570 & 85.2 & & 0.311 & 1.486 & 0.448 & 14.8 \\
V711 Tau (1988) & 0.880 & 0.362 & 5.063 & 1.821 & 70.5 & & 0.410 & 1.743 & 0.761 & 29.5 \\
\noalign{\smallskip}
MM Her & 0.498 & 0.076 & 3.367 & 0.272 & 46.1 & & 0.248 & 1.203 & 0.318 & 53.9 \\
       & 0.630 & 0.044 & 3.612 & 0.167 & 30.9 & & 0.274 & 1.283 & 0.374 & 69.1 \\
       & 0.745 & 0.064 & 3.222 & 0.221 & 44.7 & & 0.217 & 1.180 & 0.273 & 55.3 \\
\noalign{\smallskip}
V815 Her (1995) & 0.978 & 0.142 & 4.622 & 0.687 & 53.7 & & 0.397 & 1.377 & 0.582 & 45.5 \\
                & 0.520 & 0.134 & 3.577 & 0.509 & 58.2 & & 0.267 & 1.288 & 0.366 & 41.8 \\
                & 0.099 & 0.113 & 4.775 & 0.561 & 51.2 & & 0.334 & 1.467 & 0.521 & 47.6 \\
V815 Her (1989) & 0.520 & 0.092 & 4.420 & 0.416 & 60.1 & & 0.200 & 1.294 & 0.276 & 39.9 \\
\noalign{\smallskip}
HK Lac (1995) & 0.067 & 0.089 & 4.890 & 0.458 & 59.9 & & 0.291 & 0.991 & 0.307 & 40.1 \\
              & 0.110 & 0.095 & 5.286 & 0.515 & 58.1 & & 0.329 & 1.065 & 0.373 & 42.0 \\
              & 0.149 & 0.089 & 5.951 & 0.522 & 56.2 & & 0.339 & 1.129 & 0.408 & 43.9 \\
HK Lac (1989) & 0.100 & 0.142 & 5.654 & 0.726 & 58.8 & & 0.346 & 1.384 & 0.509 & 41.2 \\ 
\noalign{\smallskip}
II Peg & 0.575 & 0.211 & 3.379 & 0.758 & 35.6 & & 0.952 & 1.351 & 1.369 & 64.4 \\
       & 0.587 & 0.247 & 3.352 & 0.882 & 41.1 & & 0.913 & 1.301 & 1.264 & 58.9 \\
       & 0.735 & 0.271 & 3.880 & 1.118 & 47.3 & & 0.856 & 1.366 & 1.244 & 52.7 \\
       & 0.749 & 0.210 & 5.000 & 1.119 & 45.0 & & 0.898 & 1.430 & 1.368 & 55.0 \\
       & 0.760 & 0.245 & {\bf 7.984}& 2.077 & 56.3 & & 0.952 & 1.594 & 1.616 & 43.7 \\
       & 0.874 & 0.223 & 3.697 & 0.876 & 44.6 & & 0.750 & 1.359 & 1.086 & 55.3 \\
       & 0.890 & 0.197 & 4.049 & 0.849 & 43.1 & & 0.761 & 1.386 & 1.121 & 56.8 \\
       & 0.907 & 0.242 & 3.563 & 0.918 & 46.8 & & 0.743 & 1.319 & 1.042 & 53.1 \\
\noalign{\smallskip}
\hline
\noalign{\smallskip}
%
%
DM UMa & 0.400 & 0.207 & 10.36 & 0.810 & 27.8 & & 0.999 & 1.971 & 2.107 & 72.2 \\
       & 0.530 & 0.246 & 8.607 & 1.135 & 38.0 & & 0.918 & 1.893 & 1.850 & 62.0 \\
\noalign{\smallskip}
VV Mon & 0.710 & 0.177 & 5.382 & 0.918 & 68.2 & & 0.252 & 1.596 & 0.428 & 31.8 \\
\noalign{\smallskip}
\hline
\noalign{\smallskip}
\end{tabular}

\end{flushleft}
\end{table*}

\subsubsection{The He~{\sc i} D$_{3}$ line}

The He~{\sc i} D$_{3}$ $\lambda$5876 and He~{\sc i} $\lambda$10830 
triplets are known to be activity indicators in 
the Sun and late type stars (Zirin 1988, Shcherbakov et al. 1996). 
In the Sun, the He~{\sc i} D$_{3}$ line appears like 
an absorption feature cospatial with plages (Landman 1981), 
and it almost disappears when we look at the solar disk. This feature 
is also seen in absorption in surges, eruptive prominences, and weaker 
flares, whereas in emission in more intense flares (Zirin 1988).


The He~{\sc i} D$_{3}$ line is formed at middle chromosphere, and its
 correlation with 
X-ray flux and Ca ~{\sc ii} H \& K lines suggests that the fractional area of 
the stellar disk covered by plages may be a key factor in the formation of 
D$_{3}$ (Danks \& Lambert 1985). Moreover it has been observed a slight 
rotational modulation in $\kappa$ Cet (Lambert \& O'Brien 1983) 
and $\chi$$^{1}$ Ori (Danks \& Lambert 1985).

Historically, there have been basically two models to explain 
the line formation of He~{\sc i} D$_{3}$: 
(i) Zirin (1975) suggested that He~{\sc i} triplet levels 
were populated by over-photoionisation of the He~{\sc i} atoms by EUV 
and X-ray radiation, and subsequent radiative recombinations and cascade. 
(ii) Wolff et al. (1985) argued that collisional excitation and ionization 
in the chromosphere contributed also to the He~{\sc i} D$_{3}$ 
formation, and not only the EUV and X-ray radiation from the corona.
However, the most recent models (Andretta \& Giampapa 1995, 
Lanzafame \& Byrne 1995) 
seem to indicate that the primary mechanism in the formation of the  
He~{\sc i} triplets is the collisional excitation and ionization 
(followed by recombination cascade) by electron impact. 

The He~{\sc i} D$_{3}$ line usually appears, in stars,
 in absorption, but sometimes is in 
emission. There are two possible reasons: 
(i) Temperature and/or electronic density conditions are higher than ordinary, 
like may occur in flares (Zirin 1988, Andretta \& Giampapa 1995, Lanzafame \& 
Byrne 1995). 
(ii) As it has been seen in He~{\sc i} $\lambda$10830, 
depending on the position of the emitting region in the disk or off the limb, 
the He~{\sc i} D$_{3}$ line would appear in absorption or emission. 
Since the He~{\sc i} $\lambda$10830 is formed in emitting 
regions located at some distance from the stellar photosphere, when the 
emitting region is seen in projection against the stellar disk, He~{\sc i} 
$\lambda$10830 line appears in absorption, 
and when the emitting region is observed 
off the stellar limb, the line is in emission (Simon et al. 1982, Wolff \& 
Heasley 1984). These conclusions could extend to the case of 
He~{\sc i} D$_{3}$, since it is produced at the same region 
that He~{\sc i} $\lambda$10830. 

The He~{\sc i} D$_{3}$ line has been studied only in some
chromospherically active binaries as 
II Peg (Huenemoerder \& Ramsey 1987; Huenemoerder et al. 1990), 
DM UMa (Hatzes 1995), ER Vul (Gunn \& Doyle 1996) and  
GK Hya (Gunn et al. 1996).
The observation of emission in the He~{\sc i} D$_{3}$ line supports 
the detection of flare like events as in the case of II Peg 
(Huenemoerder \& Ramsey 1987) 
and the weak-lined T Tauri star V410 Tau (Welty \& Ramsey 1996).

In our spectra the He~{\sc i} D$_{3}$ line has been found 
in emission only during the flares of UX Ari and II Peg.
We wish to emphasize that
the detection of He~{\sc i} D$_{3}$ in emission 
in the RS CVn systems seems to occur at orbital phases near to the quadrature.
In our observations we have detected He~{\sc i} D$_{3}$ in emission
at orbital phase 0.74 in UX Ari (Montes et al. 1996b)
and at 0.76 in II Peg.
This line has been also observed in emission
at orbital phases 0.22, 0.26, 0.77 in II~Peg
by Huenemoerder \& Ramsey (1987) and  Huenemoerder et al. (1990).
 Probably we are observing a flare off the limb, 
i.e. when the plage regions are 
near the limb (the active regions are preferably in the opposite faces of the
stars), which is the most favourable situation to see an off the limb flare. 
But we cannot distinguish whether the emission is only due to 
the existence of the flare, 
or it is favoured by the relative position on the star. 

The application of the spectral subtraction to our sample
reveals that the He~{\sc i} D$_{3}$ line appears 
as an absorption feature more frequently in giants than in dwarfs.
Three out of five giants observed show clear
absorptions (BD Cet, V1149 Ori and HK Lac) and two of them exhibit
 a possible absorption (AY Cet and XX Tri), 
while among IV and V luminosity class stars there are only 
two plain absorptions.
Various authors seem to point out a more frequent presence of
He~{\sc i} $\lambda$10830 and $\lambda$5876 triplets
in giants and supergiants than in dwarfs (Simon et al. 1982,
Zirin 1982, Wolff \& Heasley 1984).

Zirin (1982) observed He~{\sc i} $\lambda$10830 usually in absorption, but
sometimes it appears in emission, especially in giant and supergiants,
with a P Cygni form, and he attributes it to
a mass-ejection phenomenon (see also O'Brien \& Lambert 1986).
Simon et al. (1982) saw that none of the single red giants, in their sample,
having strong $\lambda$10830 absorption or emission has prominent transition
region emission lines or soft X-ray emission, and they proposed a scattering
process-like responsible for the $\lambda$10830 line formation.
Smith (1983) attributed a larger intensity in $\lambda$10830 line
for giants and supergiants to the most efficient ionization
by EUV and X-ray radiation in atmospheres of coronally active giants.
Other authors say that $\lambda$10830 line is sometimes produced
 by the propagation of acoustic shock waves,
or that He~{\sc i} $\lambda$10830
transition represents a wind diagnostic.
Some of the above proposed mechanisms could also be applied
to the He~{\sc i} D$_{3}$ line.

\subsubsection{The Na~{\sc i} D$_{1}$ and D$_{2}$ lines}
                         
The Na~{\sc i} D$_{1}$ and D$_{2}$ lines
are collisionally-controlled in the
atmospheres of late-type stars and are formed in the lower chromosphere.
So, the detection of filled~-~in absorption in the D$_{1}$ and D$_{2}$ lines
may provide information about chromospheric emission.
(see the recent models of these lines 
for M dwarfs stars by Andretta et al. 1997).

In the Sun, Barrado et al. (1995) and Barrado (1996) have found 
changes in the EW of Na~{\sc i} lines  in spectra 
taken at different regions over 
the solar surface, and a relation with the filled-in absorption H$\alpha$ that 
might indicate that there is a non-radiative effect in the formation of these 
lines.

In other stars the D$_{1}$ and D$_{2}$ lines
have been observed in emission or as a filled-in
in very active red dwarf flare stars (Pettersen et al. 1984; Pettersen 1989).
However, no systematic study of these lines has been performed in stars
with different levels of activity, and in chromospherically active binaries
only the negative and uncertain detection of filled-in 
in the few active binaries ER Vul and GK Hya, respectively, 
has been reported in the recent studies 
of Gunn \& Doyle (1996) and  Gunn et al. (1996).

The application of the spectral subtraction technique in these lines is 
more difficult that in the H$\alpha$ line, 
because their wings are very sensitive to 
the effective temperature, mainly in latter spectral types.
Therefore, small differences in spectral type, 
not appreciated in the H$\alpha$ line, 
produce significant changes of the subtracted spectra in the wings of the 
Na~{\sc i} lines.
Moreover, in this spectral region there is a large number of telluric lines,
and in the spectra of some stars interstellar Na~{\sc i} could be present.
However, the distances of the majority of the stars is lower than 50 pc and 
the effect of the interstellar Na~{\sc i} is negligible.

In spite of this problems, some conclusions can be drawn.
In the chromospherically active binaries analysed here,
the spectral subtraction reveals that the
core of the Na~{\sc i} D$_{1}$ and D$_{2}$
lines are filled-in by
chromospheric emission in the more active star of the sample
(the star with H$\alpha$ emission above the continuum, and with larger
excess H$\alpha$ emission EW).
The stars with only a small or without excess H$\alpha$ emission 
as BD Cet, AY Cet, V1149 Ori and KT Peg do not exhibit 
excess emission in the Na~{\sc i} lines.
Moreover, the excess D$_{1}$ and D$_{2}$ emissions obtained 
are larger in the systems with larger
excess H$\alpha$ emission, 
and also increase in the flares observed in UX Ari and II Peg.
In short, we can conclude that 
the filled-in of the core of 
the Na~{\sc i} D$_{1}$ and D$_{2}$ lines 
could be used as a chromospheric activity indicator.

\section{Conclusions}

In this paper we have analysed, using the spectral subtraction technique,
simultaneous H$\alpha$, Na~{\sc i} D$_{1}$, D$_{2}$,
and He~{\sc i} D$_{3}$ spectroscopic observations
of 18 chromospherically active binary systems.

We have found excess H$\alpha$ in all the systems, except KT Peg which  
have also small emission in the Ca~{\sc ii} H  \& K lines.
The subtracted H$\alpha$ profile of the more active stars of the sample 
(H$\alpha$ in emission above the continuum)   
has very broad wings, and is well matched using a two-components  
Gaussian fit (narrow and broad).
The broad component is primarily responsible for the observed variations
of the profile, and its contribution to the total EW increases with
the degree of activity.
So, we have interpreted this broad component emission
as arising from microflaring activity that take place 
in the chromosphere of this very active stars.

The H$\alpha$ observation of the eclipsing binary system AR Lac at orbital 
phase 0.939, when a 0.33 fraction of the hot component is hidden by the 
cool component, allowed us to detect the presence of prominence-like 
extended material seen off the limb of the cool component 
through the detection of excess H$\alpha$ absorption
located at the wavelength position corresponding to the hot component.
A small excess absorption in the He~{\sc i} D$_{3}$ line is also present 
at this orbital phase.

We reported the detection of an optical flare in the 
systems UX Ari and II Peg through
the presence of the He~{\sc i} D$_{3}$ in emission in coincidence with the
enhancement of the H$\alpha$ emission.

We have found the  He~{\sc i} D$_{3}$ in emission only in the two 
above mentioned flares, 
and as an absorption feature in the subtracted spectra of 
a large number of giant stars of the sample.

The application of the spectral subtraction technique reveals that the
core of the Na~{\sc i} D$_{1}$ and D$_{2}$ lines are also filled-in by
chromospheric emission in the more active star of the sample.
The stars with only a small excess H$\alpha$ emission do not exhibit 
excess emission in the Na~{\sc i} lines.



\begin{acknowledgements}

This work has been supported by the Universidad Complutense de Madrid
and the Spanish Direcci\'{o}n General de Investigaci\'{o}n
Cient\'{\i}fica y  T\'{e}cnica (DGICYT) under grant PB94-0263.
We would like to thank the referee S. Catalano
for suggesting several improvements and clarifications.

\end{acknowledgements}



        
                                                         

\end{document}